\documentclass[12pt]{iopart}

\usepackage{iopams}

\usepackage{graphicx}
\usepackage{subfig}

\usepackage[jphysicsB]{harvard}

\newcommand{\dod}[2]{\frac{\partial #1}{\partial{#2}}}
\newcommand{\dsods}[2]{\frac{\partial^2 #1}{\partial{#2}^2}}

\providecommand\bcdot{\boldsymbol{\cdot}}

\begin{document}
\def\PsfigVersion{1.9}
\ifx\undefined\psfig\else \fi

%

\let\LaTeXAtSign=\@
\let\@=\relax
\edef\psfigRestoreAt{\catcode`\@=\number\catcode`@\relax}
\catcode`\@=11\relax
\newwrite\@unused
\def\ps@typeout#1{{\let\protect\string\immediate\write\@unused{#1}}}
\ps@typeout{psfig/tex \PsfigVersion}


\def\figurepath{./}
\def\psfigurepath#1{\edef\figurepath{#1}}

%
%
\def\@nnil{\@nil}
\def\@empty{}
\def\@psdonoop#1\@@#2#3{}
\def\@psdo#1:=#2\do#3{\edef\@psdotmp{#2}\ifx\@psdotmp\@empty \else
    \expandafter\@psdoloop#2,\@nil,\@nil\@@#1{#3}\fi}
\def\@psdoloop#1,#2,#3\@@#4#5{\def#4{#1}\ifx #4\@nnil \else
       #5\def#4{#2}\ifx #4\@nnil \else#5\@ipsdoloop #3\@@#4{#5}\fi\fi}
\def\@ipsdoloop#1,#2\@@#3#4{\def#3{#1}\ifx #3\@nnil 
       \let\@nextwhile=\@psdonoop \else
      #4\relax\let\@nextwhile=\@ipsdoloop\fi\@nextwhile#2\@@#3{#4}}
\def\@tpsdo#1:=#2\do#3{\xdef\@psdotmp{#2}\ifx\@psdotmp\@empty \else
    \@tpsdoloop#2\@nil\@nil\@@#1{#3}\fi}
\def\@tpsdoloop#1#2\@@#3#4{\def#3{#1}\ifx #3\@nnil 
       \let\@nextwhile=\@psdonoop \else
      #4\relax\let\@nextwhile=\@tpsdoloop\fi\@nextwhile#2\@@#3{#4}}
%
\ifx\undefined\fbox
\newdimen\fboxrule
\newdimen\fboxsep
\newdimen\ps@tempdima
\newbox\ps@tempboxa
\fboxsep = 3pt
\fboxrule = .4pt
\long\def\fbox#1{\leavevmode\setbox\ps@tempboxa\hbox{#1}\ps@tempdima\fboxrule
    \advance\ps@tempdima \fboxsep \advance\ps@tempdima \dp\ps@tempboxa
   \hbox{\lower \ps@tempdima\hbox
  {\vbox{\hrule height \fboxrule
          \hbox{\vrule width \fboxrule \hskip\fboxsep
          \vbox{\vskip\fboxsep \box\ps@tempboxa\vskip\fboxsep}\hskip 
                 \fboxsep\vrule width \fboxrule}
                 \hrule height \fboxrule}}}}
\fi
%
%
\newread\ps@stream
\newif\ifnot@eof       
\newif\if@noisy        
\newif\if@atend        
\newif\if@psfile       
%
%
{\catcode`\%=12\global\gdef\epsf@start{
\def\epsf@PS{PS}
\def\epsf@getbb#1{%
%
%
\openin\ps@stream=#1
\ifeof\ps@stream\ps@typeout{Error, File #1 not found}\else
%
%
   {\not@eoftrue \chardef\other=12
    \def\do##1{\catcode`##1=\other}\dospecials \catcode`\ =10
    \loop
       \if@psfile
	  \read\ps@stream to \epsf@fileline
       \else{
	  \obeyspaces
          \read\ps@stream to \epsf@tmp\global\let\epsf@fileline\epsf@tmp}
       \fi
       \ifeof\ps@stream\not@eoffalse\else
%
%
       \if@psfile\else
       \expandafter\epsf@test\epsf@fileline:. \\%
       \fi
%
%
          \expandafter\epsf@aux\epsf@fileline:. \\%
       \fi
   \ifnot@eof\repeat
   }\closein\ps@stream\fi}%
%
%
\long\def\epsf@test#1#2#3:#4\\{\def\epsf@testit{#1#2}
			\ifx\epsf@testit\epsf@start\else
\ps@typeout{Warning! File does not start with `\epsf@start'.  It may not be a PostScript file.}
			\fi
			\@psfiletrue} 
%
%
{\catcode`\%=12\global\let\epsf@percent=
%
%
%
\long\def\epsf@aux#1#2:#3\\{\ifx#1\epsf@percent
   \def\epsf@testit{#2}\ifx\epsf@testit\epsf@bblit
	\@atendfalse
        \epsf@atend #3 . \\%
	\if@atend	
	   \if@verbose{
		\ps@typeout{psfig: found `(atend)'; continuing search}
	   }\fi
        \else
        \epsf@grab #3 . . . \\%
        \not@eoffalse
        \global\no@bbfalse
        \fi
   \fi\fi}%
%
%
\def\epsf@grab #1 #2 #3 #4 #5\\{%
   \global\def\epsf@llx{#1}\ifx\epsf@llx\empty
      \epsf@grab #2 #3 #4 #5 .\\\else
   \global\def\epsf@lly{#2}%
   \global\def\epsf@urx{#3}\global\def\epsf@ury{#4}\fi}%
%
%
\def\epsf@atendlit{(atend)} 
\def\epsf@atend #1 #2 #3\\{%
   \def\epsf@tmp{#1}\ifx\epsf@tmp\empty
      \epsf@atend #2 #3 .\\\else
   \ifx\epsf@tmp\epsf@atendlit\@atendtrue\fi\fi}


\chardef\psletter = 11 
\chardef\other = 12

\newif \ifdebug 
\newif\ifc@mpute 
\c@mputetrue 

\let\then = \relax
\def\r@dian{pt }
\let\r@dians = \r@dian
\let\dimensionless@nit = \r@dian
\let\dimensionless@nits = \dimensionless@nit
\def\internal@nit{sp }
\let\internal@nits = \internal@nit
\newif\ifstillc@nverging
\def \Mess@ge #1{\ifdebug \then \message {#1} \fi}

{ 
	\catcode `\@ = \psletter
	\gdef \nodimen {\expandafter \n@dimen \the \dimen}
	\gdef \term #1 #2 #3%
	       {\edef \t@ {\the #1}
		\edef \t@@ {\expandafter \n@dimen \the #2\r@dian}%
		\t@rm {\t@} {\t@@} {#3}%
	       }
	\gdef \t@rm #1 #2 #3%
	       {{%
		\count 0 = 0
		\dimen 0 = 1 \dimensionless@nit
		\dimen 2 = #2\relax
		\Mess@ge {Calculating term #1 of \nodimen 2}%
		\loop
		\ifnum	\count 0 < #1
		\then	\advance \count 0 by 1
			\Mess@ge {Iteration \the \count 0 \space}%
			\Multiply \dimen 0 by {\dimen 2}%
			\Mess@ge {After multiplication, term = \nodimen 0}%
			\Divide \dimen 0 by {\count 0}%
			\Mess@ge {After division, term = \nodimen 0}%
		\repeat
		\Mess@ge {Final value for term #1 of 
				\nodimen 2 \space is \nodimen 0}%
		\xdef \Term {#3 = \nodimen 0 \r@dians}%
		\aftergroup \Term
	       }}
	\catcode `\p = \other
	\catcode `\t = \other
	\gdef \n@dimen #1pt{#1} 
}

\def \Divide #1by #2{\divide #1 by #2} 

\def \Multiply #1by #2
       {{
	\count 0 = #1\relax
	\count 2 = #2\relax
	\count 4 = 65536
	\Mess@ge {Before scaling, count 0 = \the \count 0 \space and
			count 2 = \the \count 2}%
	\ifnum	\count 0 > 32767 
	\then	\divide \count 0 by 4
		\divide \count 4 by 4
	\else	\ifnum	\count 0 < -32767
		\then	\divide \count 0 by 4
			\divide \count 4 by 4
		\else
		\fi
	\fi
	\ifnum	\count 2 > 32767 
	\then	\divide \count 2 by 4
		\divide \count 4 by 4
	\else	\ifnum	\count 2 < -32767
		\then	\divide \count 2 by 4
			\divide \count 4 by 4
		\else
		\fi
	\fi
	\multiply \count 0 by \count 2
	\divide \count 0 by \count 4
	\xdef \product {#1 = \the \count 0 \internal@nits}%
	\aftergroup \product
       }}

\def\r@duce{\ifdim\dimen0 > 90\r@dian \then   
		\multiply\dimen0 by -1
		\advance\dimen0 by 180\r@dian
		\r@duce
	    \else \ifdim\dimen0 < -90\r@dian \then  
		\advance\dimen0 by 360\r@dian
		\r@duce
		\fi
	    \fi}

\def\Sine#1%
       {{%
	\dimen 0 = #1 \r@dian
	\r@duce
	\ifdim\dimen0 = -90\r@dian \then
	   \dimen4 = -1\r@dian
	   \c@mputefalse
	\fi
	\ifdim\dimen0 = 90\r@dian \then
	   \dimen4 = 1\r@dian
	   \c@mputefalse
	\fi
	\ifdim\dimen0 = 0\r@dian \then
	   \dimen4 = 0\r@dian
	   \c@mputefalse
	\fi
	\ifc@mpute \then
		\divide\dimen0 by 180
		\dimen0=3.141592654\dimen0
		\dimen 2 = 3.1415926535897963\r@dian 
		\divide\dimen 2 by 2 
		\Mess@ge {Sin: calculating Sin of \nodimen 0}%
		\count 0 = 1 
		\dimen 2 = 1 \r@dian 
		\dimen 4 = 0 \r@dian 
		\loop
			\ifnum	\dimen 2 = 0 
			\then	\stillc@nvergingfalse 
			\else	\stillc@nvergingtrue
			\fi
			\ifstillc@nverging 
			\then	\term {\count 0} {\dimen 0} {\dimen 2}%
				\advance \count 0 by 2
				\count 2 = \count 0
				\divide \count 2 by 2
				\ifodd	\count 2 
				\then	\advance \dimen 4 by \dimen 2
				\else	\advance \dimen 4 by -\dimen 2
				\fi
		\repeat
	\fi		
			\xdef \sine {\nodimen 4}%
       }}

\def\Cosine#1{\ifx\sine\UnDefined\edef\Savesine{\relax}\else
		             \edef\Savesine{\sine}\fi
	{\dimen0=#1\r@dian\advance\dimen0 by 90\r@dian
	 \Sine{\nodimen 0}
	 \xdef\cosine{\sine}
	 \xdef\sine{\Savesine}}}	      

\def\psdraft{
	\def\@psdraft{0}
}
\def\psfull{
	\def\@psdraft{100}
}

\psfull

\newif\if@scalefirst
\def\psscalefirst{\@scalefirsttrue}
\def\psrotatefirst{\@scalefirstfalse}
\psrotatefirst

\newif\if@draftbox
\def\psnodraftbox{
	\@draftboxfalse
}
\def\psdraftbox{
	\@draftboxtrue
}
\@draftboxtrue

\newif\if@prologfile
\newif\if@postlogfile
\def\pssilent{
	\@noisyfalse
}
\def\psnoisy{
	\@noisytrue
}
\psnoisy
\newif\if@bbllx
\newif\if@bblly
\newif\if@bburx
\newif\if@bbury
\newif\if@height
\newif\if@width
\newif\if@rheight
\newif\if@rwidth
\newif\if@angle
\newif\if@clip
\newif\if@verbose
\def\@p@@sclip#1{\@cliptrue}

\newif\if@decmpr


\def\@p@@sfigure#1{\def\@p@sfile{null}\def\@p@sbbfile{null}
	        \openin1=#1.bb
		\ifeof1\closein1
	        	\openin1=\figurepath#1.bb
			\ifeof1\closein1
			        \openin1=#1
				\ifeof1\closein1%
				       \openin1=\figurepath#1
					\ifeof1
					   \ps@typeout{Error, File #1 not found}
						\if@bbllx\if@bblly
				   		\if@bburx\if@bbury
			      				\def\@p@sfile{#1}%
			      				\def\@p@sbbfile{#1}%
							\@decmprfalse
				  	   	\fi\fi\fi\fi
					\else\closein1
				    		\def\@p@sfile{\figurepath#1}%
				    		\def\@p@sbbfile{\figurepath#1}%
						\@decmprfalse
	                       		\fi%
			 	\else\closein1%
					\def\@p@sfile{#1}
					\def\@p@sbbfile{#1}
					\@decmprfalse
			 	\fi
			\else
				\def\@p@sfile{\figurepath#1}
				\def\@p@sbbfile{\figurepath#1.bb}
				\@decmprtrue
			\fi
		\else
			\def\@p@sfile{#1}
			\def\@p@sbbfile{#1.bb}
			\@decmprtrue
		\fi}

\def\@p@@sfile#1{\@p@@sfigure{#1}}

\def\@p@@sbbllx#1{
		\@bbllxtrue
		\dimen100=#1
		\edef\@p@sbbllx{\number\dimen100}
}
\def\@p@@sbblly#1{
		\@bbllytrue
		\dimen100=#1
		\edef\@p@sbblly{\number\dimen100}
}
\def\@p@@sbburx#1{
		\@bburxtrue
		\dimen100=#1
		\edef\@p@sbburx{\number\dimen100}
}
\def\@p@@sbbury#1{
		\@bburytrue
		\dimen100=#1
		\edef\@p@sbbury{\number\dimen100}
}
\def\@p@@sheight#1{
		\@heighttrue
		\dimen100=#1
   		\edef\@p@sheight{\number\dimen100}
}
\def\@p@@swidth#1{
		\@widthtrue
		\dimen100=#1
		\edef\@p@swidth{\number\dimen100}
}
\def\@p@@srheight#1{
		\@rheighttrue
		\dimen100=#1
		\edef\@p@srheight{\number\dimen100}
}
\def\@p@@srwidth#1{
		\@rwidthtrue
		\dimen100=#1
		\edef\@p@srwidth{\number\dimen100}
}
\def\@p@@sangle#1{
		\@angletrue
		\edef\@p@sangle{#1} 
}
\def\@p@@ssilent#1{ 
		\@verbosefalse
}
\def\@p@@sprolog#1{\@prologfiletrue\def\@prologfileval{#1}}
\def\@p@@spostlog#1{\@postlogfiletrue\def\@postlogfileval{#1}}
\def\@cs@name#1{\csname #1\endcsname}
\def\@setparms#1=#2,{\@cs@name{@p@@s#1}{#2}}
%
%
\def\ps@init@parms{
		\@bbllxfalse \@bbllyfalse
		\@bburxfalse \@bburyfalse
		\@heightfalse \@widthfalse
		\@rheightfalse \@rwidthfalse
		\def\@p@sbbllx{}\def\@p@sbblly{}
		\def\@p@sbburx{}\def\@p@sbbury{}
		\def\@p@sheight{}\def\@p@swidth{}
		\def\@p@srheight{}\def\@p@srwidth{}
		\def\@p@sangle{0}
		\def\@p@sfile{} \def\@p@sbbfile{}
		\def\@p@scost{10}
		\def\@sc{}
		\@prologfilefalse
		\@postlogfilefalse
		\@clipfalse
		\if@noisy
			\@verbosetrue
		\else
			\@verbosefalse
		\fi
}
%
%
\def\parse@ps@parms#1{
	 	\@psdo\@psfiga:=#1\do
		   {\expandafter\@setparms\@psfiga,}}
%
%
\newif\ifno@bb
\def\bb@missing{
	\if@verbose{
		\ps@typeout{psfig: searching \@p@sbbfile \space  for bounding box}
	}\fi
	\no@bbtrue
	\epsf@getbb{\@p@sbbfile}
        \ifno@bb \else \bb@cull\epsf@llx\epsf@lly\epsf@urx\epsf@ury\fi
}	
\def\bb@cull#1#2#3#4{
	\dimen100=#1 bp\edef\@p@sbbllx{\number\dimen100}
	\dimen100=#2 bp\edef\@p@sbblly{\number\dimen100}
	\dimen100=#3 bp\edef\@p@sbburx{\number\dimen100}
	\dimen100=#4 bp\edef\@p@sbbury{\number\dimen100}
	\no@bbfalse
}
\newdimen\p@intvaluex
\newdimen\p@intvaluey
\def\rotate@#1#2{{\dimen0=#1 sp\dimen1=#2 sp
		  \global\p@intvaluex=\cosine\dimen0
		  \dimen3=\sine\dimen1
		  \global\advance\p@intvaluex by -\dimen3
		  \global\p@intvaluey=\sine\dimen0
		  \dimen3=\cosine\dimen1
		  \global\advance\p@intvaluey by \dimen3
		  }}
\def\compute@bb{
		\no@bbfalse
		\if@bbllx \else \no@bbtrue \fi
		\if@bblly \else \no@bbtrue \fi
		\if@bburx \else \no@bbtrue \fi
		\if@bbury \else \no@bbtrue \fi
		\ifno@bb \bb@missing \fi
		\ifno@bb \ps@typeout{FATAL ERROR: no bb supplied or found}
			\no-bb-error
		\fi
		%
%
		\count203=\@p@sbburx
		\count204=\@p@sbbury
		\advance\count203 by -\@p@sbbllx
		\advance\count204 by -\@p@sbblly
		\edef\ps@bbw{\number\count203}
		\edef\ps@bbh{\number\count204}
		\if@angle 
			\Sine{\@p@sangle}\Cosine{\@p@sangle}
	        	{\dimen100=\maxdimen\xdef\r@p@sbbllx{\number\dimen100}
					    \xdef\r@p@sbblly{\number\dimen100}
			                    \xdef\r@p@sbburx{-\number\dimen100}
					    \xdef\r@p@sbbury{-\number\dimen100}}
%
                        \def\minmaxtest{
			   \ifnum\number\p@intvaluex<\r@p@sbbllx
			      \xdef\r@p@sbbllx{\number\p@intvaluex}\fi
			   \ifnum\number\p@intvaluex>\r@p@sbburx
			      \xdef\r@p@sbburx{\number\p@intvaluex}\fi
			   \ifnum\number\p@intvaluey<\r@p@sbblly
			      \xdef\r@p@sbblly{\number\p@intvaluey}\fi
			   \ifnum\number\p@intvaluey>\r@p@sbbury
			      \xdef\r@p@sbbury{\number\p@intvaluey}\fi
			   }
			\rotate@{\@p@sbbllx}{\@p@sbblly}
			\minmaxtest
			\rotate@{\@p@sbbllx}{\@p@sbbury}
			\minmaxtest
			\rotate@{\@p@sbburx}{\@p@sbblly}
			\minmaxtest
			\rotate@{\@p@sbburx}{\@p@sbbury}
			\minmaxtest
			\edef\@p@sbbllx{\r@p@sbbllx}\edef\@p@sbblly{\r@p@sbblly}
			\edef\@p@sbburx{\r@p@sbburx}\edef\@p@sbbury{\r@p@sbbury}
		\fi
		\count203=\@p@sbburx
		\count204=\@p@sbbury
		\advance\count203 by -\@p@sbbllx
		\advance\count204 by -\@p@sbblly
		\edef\@bbw{\number\count203}
		\edef\@bbh{\number\count204}
}
%
%
\def\in@hundreds#1#2#3{\count240=#2 \count241=#3
		     \count100=\count240	
		     \divide\count100 by \count241
		     \count101=\count100
		     \multiply\count101 by \count241
		     \advance\count240 by -\count101
		     \multiply\count240 by 10
		     \count101=\count240	
		     \divide\count101 by \count241
		     \count102=\count101
		     \multiply\count102 by \count241
		     \advance\count240 by -\count102
		     \multiply\count240 by 10
		     \count102=\count240	
		     \divide\count102 by \count241
		     \count200=#1\count205=0
		     \count201=\count200
			\multiply\count201 by \count100
		 	\advance\count205 by \count201
		     \count201=\count200
			\divide\count201 by 10
			\multiply\count201 by \count101
			\advance\count205 by \count201
		     \count201=\count200
			\divide\count201 by 100
			\multiply\count201 by \count102
			\advance\count205 by \count201
		     \edef\@result{\number\count205}
}
\def\compute@wfromh{
		\in@hundreds{\@p@sheight}{\@bbw}{\@bbh}
		\edef\@p@swidth{\@result}
}
\def\compute@hfromw{
	        \in@hundreds{\@p@swidth}{\@bbh}{\@bbw}
		\edef\@p@sheight{\@result}
}
\def\compute@handw{
		\if@height 
			\if@width
			\else
				\compute@wfromh
			\fi
		\else 
			\if@width
				\compute@hfromw
			\else
				\edef\@p@sheight{\@bbh}
				\edef\@p@swidth{\@bbw}
			\fi
		\fi
}
\def\compute@resv{
		\if@rheight \else \edef\@p@srheight{\@p@sheight} \fi
		\if@rwidth \else \edef\@p@srwidth{\@p@swidth} \fi
}
%
\def\compute@sizes{
	\compute@bb
	\if@scalefirst\if@angle
	\if@width
	   \in@hundreds{\@p@swidth}{\@bbw}{\ps@bbw}
	   \edef\@p@swidth{\@result}
	\fi
	\if@height
	   \in@hundreds{\@p@sheight}{\@bbh}{\ps@bbh}
	   \edef\@p@sheight{\@result}
	\fi
	\fi\fi
	\compute@handw
	\compute@resv}

%
%
\def\psfig#1{\vbox {
	%
	\ps@init@parms
	\parse@ps@parms{#1}
	\compute@sizes
	\ifnum\@p@scost<\@psdraft{
		\special{ps::[begin] 	\@p@swidth \space \@p@sheight \space
				\@p@sbbllx \space \@p@sbblly \space
				\@p@sbburx \space \@p@sbbury \space
				startTexFig \space }
		\if@angle
			\special {ps:: \@p@sangle \space rotate \space} 
		\fi
		\if@clip{
			\if@verbose{
				\ps@typeout{(clip)}
			}\fi
			\special{ps:: doclip \space }
		}\fi
		\if@prologfile
		    \special{ps: plotfile \@prologfileval \space } \fi
		\if@decmpr{
			\if@verbose{
				\ps@typeout{psfig: including \@p@sfile.Z \space }
			}\fi
			\special{ps: plotfile "`zcat \@p@sfile.Z" \space }
		}\else{
			\if@verbose{
				\ps@typeout{psfig: including \@p@sfile \space }
			}\fi
			\special{ps: plotfile \@p@sfile \space }
		}\fi
		\if@postlogfile
		    \special{ps: plotfile \@postlogfileval \space } \fi
		\special{ps::[end] endTexFig \space }
		\vbox to \@p@srheight sp{
			\hbox to \@p@srwidth sp{
				\hss
			}
		\vss
		}
	}\else{
		\if@draftbox{		
			\hbox{\frame{\vbox to \@p@srheight sp{
			\vss
			\hbox to \@p@srwidth sp{ \hss \@p@sfile \hss }
			\vss
			}}}
		}\else{
			\vbox to \@p@srheight sp{
			\vss
			\hbox to \@p@srwidth sp{\hss}
			\vss
			}
		}\fi

	}\fi
}}
\psfigRestoreAt
\let\@=\LaTeXAtSign


\article[The Influence of Horizontal Boundaries on EC and AMT in a
Cylindrical Annulus]{ First International Conference ``Turbulent
Mixing and Beyond''\footnote{held on 18-26 of August 2007 at the
Abdus Salam International Centre for Theoretical Physics, Trieste,
Italy} }{The Influence of Horizontal Boundaries on Ekman Circulation
and Angular Momentum Transport in a Cylindrical Annulus}


\author{Aleksandr V. Obabko$^1$, Fausto Cattaneo$^{1,2}$ and Paul F. Fischer$^2$}

\address{$^1$Department of Astronomy and Astrophysics,
University of Chicago, Chicago, IL 60637, USA}
\address{$^2$ Division of
Mathematics and Computer Science, Argonne National Laboratory,
Argonne, IL 60439, USA} \ead{obabko@uchicago.edu}

\begin{abstract}

We present numerical simulations of circular Couette flow in
axisymmetric and fully three-dimensional geometry of a cylindrical
annulus inspired by Princeton MRI liquid gallium experiment.  The
incompressible Navier-Stokes equations are solved with the spectral
element code Nek5000 incorporating realistic horizontal boundary
conditions of differentially rotating rings.  We investigate the
effect of changing rotation rates (Reynolds number) and of the
horizontal boundary conditions on flow structure, Ekman circulation
and associated transport of angular momentum through the onset of
unsteadiness and three-dimensionality.  A mechanism for the
explanation of the dependence of the Ekman flows and circulation on
horizontal boundary conditions is proposed.

\end{abstract}

\vspace{2pc} \noindent{\it Keywords}: Navier-Stokes equations,
circular Couette flow, Ekman flow, Ekman circulation, Ekman boundary
layer, angular momentum transport, spectral element method
\maketitle

\section{Introduction}
\label{s:intro}

The phenomenon of Ekman circulation (EC) occurs in most if not all
rotating flows with stressed boundaries that are not parallel to the
axis of rotation. The manifestation of EC ranges from wind-driven
ocean currents \cite{Ba67}, to the accumulation of the tea leaves at
the bottom of a stirred cup \citeaffixed{AH60}{see, e.g.,}. One of
consequences of EC and of the associated Ekman flows is greatly to
to enhance mixing and transport and in particular, the transport of
angular momentum, above the values due to viscosity alone.
Traditionally, Ekman flows are explained in terms of action of
Coriolis forces in the Ekman layers along the rotating stressed
boundaries \cite{Gr68}.

There are circumstances when the presence of EC has undesirable
effects. For example, this is the case in laboratory experiments to
study the development of magneto-rotational instability (MRI) in
liquid metals \citeaffixed{RRB04}{see a monograph edited by}. The
MRI instability is important in astrophysics  where it is believed
to lead to turbulence in magnetized accretion disks \cite{Ba03}.
Many of the features of the MRI and its associated enhancement of
angular momentum transport (AMT) can be studied experimentally in
magnetized flows between rotating coaxial cylinders.  In these
experiments, the rotation rates of the cylinders are chosen in such
a way that the fluid's angular momentum increases outwards so that
the resulting rotational profile is stable to axisymmetric
perturbations (so-called centrifugally stable regime).  The presence
of a weak magnetic field can destabilize the basic flow, provided
the angular velocity increases inward, and lead to an enhancement of
outward AMT.

In an ideal situation, the basic state consists of circular Couette
flow (CCF), and the outward transport of angular momentum in the
absence of magnetic fields is solely due to viscous effects. The
presence of a magnetic field would destabilize the basic flow
through the effects of MRI and lead to a measurable increase of AMT.
In practice, this ideal case can never be realized in laboratory
experiments because of horizontal boundaries. The presence of these
boundaries drives an EC that enhances AMT even in the absence of
magnetic effects.  In order to study the enhancement of AMT due to
MRI it is crucial to be able to distinguish the effects that are
magnetic in origin from those that are due to the EC.  One
possibility is to make the cylinders very tall so the horizontal
boundaries are far removed from the central region.  This approach,
however, is not practical owing to the high price of liquid metals.

The alternative approach is to device boundaries in such a way that
the resulting EC can be controlled and possibly reduced.  For
example, attaching the horizontal boundaries to the inner or outer
cylinder results in dramatically different flow patterns.  Another
possibility could be to have the horizontal boundaries rotating
independently of inner and outer cylinder.  Goodman, Ji and
coworkers \cite{KJGCS04,BJSC06,JBSG06} have proposed to split the
horizontal boundaries into two independently rotating rings whose
rotational speeds are chosen so as to minimize the disruption to the
basic CCF by secondary Ekman circulations.  Indeed this approach has
been implemented in the Princeton's MRI liquid gallium experiment
\cite{Sc08}.  In any case, no matter how the horizontal boundaries
are implemented it is important to understand what kind of EC
patterns arise before the magnetic effects are introduced.

In the present paper we address this issue by studying the effects
of horizontal boundary conditions on CCF numerically.  We study both
axisymmetric and fully three-dimensional geometries and investigate
the effects of changing rotation rates (Reynolds number) through the
onset of unsteadiness and three-dimensionality. The next section
(section~\ref{s:formul}) describes the formulation of the problem
and gives an account of numerical aspects of its solution technique
including a brief description of the spectral element code Nek5000
\cite{FOC08}. The section~\ref{s:result} starts with an explanation
of flow behaviour due to horizontal boundary conditions, i.e.~CCF,
Ekman and disrupted Ekman circulation due to periodic horizontal
boundaries, `lids' and `rings', correspondingly
(section~\ref{ss:BC}). Then the paper proceeds with description of
comparison of our results with the experimental data
(section~\ref{ss:exp}) followed by an examination of torque and AMT
(section~\ref{ss:amf}). Finally, we draw conclusions and describe
future work in section~\ref{s:concl}.

\section{Problem Formulation and Numerical Method}
\label{s:formul}

\subsection{Formulation}
\label{ss:formul}

We study the flow of an incompressible fluid with finite (constant)
kinematic viscosity $\nu$ in a cylindrical annulus bounded by
coaxial cylinders.  The cylinders have the radii $R_1^*$ and $R_2^*$
($R_1^*<R_2^*$) and rotate with angular velocities $\Omega_1^*$ and
$\Omega_2^*$, respectively.  The annulus is confined in the vertical
direction by horizontal boundaries at distance $H^*$ apart.  The
formulation of the problem in cylindrical coordinates $(r,\theta,z)$
with the scales for characteristic length $L$ and velocity $U$,
\begin{equation}        \label{e:nondim:LUB}
L = R_2^* - R_1^*  \qquad\qquad  U = \Omega_1^* R_1^* - \Omega_2^*
R_2^*
\end{equation}
and therefore, with the relationship between dimensional variables
(with asterisk) and non-dimensional radius, height, velocity vector
$\boldsymbol{V}$, time and pressure given by
\begin{equation}        \label{e:nondim}
\left[ r^*, z^*, \boldsymbol{V^*}, t^*, p^* \right] = \left[ L \, r,
L \, z, U \boldsymbol{V}, \frac{L}{U} \, t, \rho U^2 p \right]
\end{equation}
correspondingly, results in the following non-dimensional
incompressible Navier-Stokes equations:
 \begin{eqnarray}
     \dod{V_r}{t} + \left( \boldsymbol{V} \bcdot \boldsymbol{\nabla} \right)V_r
     - \frac{V_\theta^2}{r}
     & = &
     \frac{1}{Re} \left[ \triangle {V_r} - \frac{2}{r^2}\dod{u_\theta}{\theta}
     - \frac{V_r}{r^2} \right] - \dod{p}{r}   \label{e:MHD:V_r}   \\
     \dod{V_\theta}{t} + \left( \boldsymbol{V} \bcdot \boldsymbol{\nabla} \right)V_\theta
     + \frac{V_r V_\theta}{r}
     & = &
     \frac{1}{Re} \left[ \triangle {V_\theta} + \frac{2}{r^2}\dod{u_r}{\theta}
     - \frac{V_\theta}{r^2} \right] - \frac{1}{r}\dod{p}{\theta}                  \label{e:MHD:V_t}   \\
     \dod{V_z}{t} + \left( \boldsymbol{V} \bcdot \boldsymbol{\nabla} \right)V_z
     & = &
     \frac{1}{Re}        \triangle {V_z}
     - \dod{p}{z}                       \label{e:MHD:V_z}   \\
     \dod{V_r}{r} + \frac{1}{r} \dod{V_\theta}{\theta} + \dod{V_z}{z} + \frac{V_r}{r}
     & = & 0                            \label{e:MHD:div_V:axi}
  \end{eqnarray}
where $\rho$ is a constant fluid density and Reynolds number $Re$ is
defined as
\begin{equation}    \label{e:Rem}
Re = \frac{U L}{\nu} = \frac{(\Omega_1^* R_1^* - \Omega_2^* R_2^*)(
R_2^* - R_1^*)}{\nu}
\end{equation}
while the scalar advection operator due to a vector field
$\boldsymbol{V}$ and laplacian of a scalar function $S(r,z)$ are
given by
\begin{equation}    \label{e:lapl:adv:axi}
\fl\hspace{7ex} \left( \boldsymbol{V} \bcdot \boldsymbol{\nabla}
\right)S = V_r \dod{S}{r} + \frac{V_\theta}{r}\dod{S}{\theta} + V_z
\dod{S}{z} \quad \triangle S = \dsods{S}{r} + \frac{1}{r}\dod{S}{r}
+ \frac{1}{r^2}\dsods{S}{\theta} + \dsods{S}{z}
\end{equation}



The initial conditions for the flow in the annulus and boundary
conditions at the cylinder surfaces $r = R_1$ and $r = R_2$ are
\begin{equation}    \label{e:BC}
V_r = V_z = 0 \qquad V_\theta = r \, \Omega(r)
\end{equation}
where non-dimensional angular velocity $\Omega(r)$ is given by
circular Couette flow (CCF) profile
\begin{equation}    \label{e:BC:OC}
%
\Omega_C(r) = A + \frac{B}{r^2} \qquad A = \frac{\Omega_2
R_2^2-\Omega_1 R_1^2}{R_2^2-R_1^2} \quad B = \frac{R_1^2
R_2^2(\Omega_1-\Omega_2)}{R_2^2-R_1^2}
\end{equation}

At the horizontal boundaries $z = 0$ and $z = H$, two types of the
boundary conditions have been considered, namely, {\it lids} and
{\it rings}, given by (\ref{e:BC}) where angular velocity
$\Omega(r)$ is equal to
\begin{equation}    \label{e:BC:O}
\Omega(r) = \left\{
\begin{array}{c@{\quad : \quad}c} \Omega_1  &  r = R_1
\\ \Omega_3  &  R_1 < r < R_{12}
\\ \Omega_4  &  R_{12} < r < R_2
\\ \Omega_2  &  r = R_2
\end{array}
\right.
\end{equation}
Here $R_{12}$ is the radial location of the boundary between the
inner and outer rings, and $\Omega_3$ and $\Omega_4$ are angular
velocities of inner and outer rings, correspondingly.  Inspired by
Princeton MRI liquid gallium experiment \cite{Sc08}, the
non-dimensional angular velocities and cylinder height as well as
cylinder and ring boundary radii used in this study are given in
table~\ref{t:param} in addition to the dimensional parameters
involved in comparison with the experiment
(subsection~\ref{ss:exp}).  In the cases with lids, angular
velocities $\Omega_3$ and $\Omega_4$ are equal to the angular
velocity of the outer cylinder $\Omega_2$ while in the cases with
rings they turn out to be close to the values of CCF profile
(\ref{e:BC:OC}) taken at the middle of radii of the corresponding
rings.

\begin{table}
  \centering
  \begin{minipage}[c]{0.4\textwidth}
    \centering
    {\renewcommand{\arraystretch}{1.3}
\begin{tabular}{c|cc||cc|c}
            & Lids   & Rings & \multicolumn{3}{c}{Experiment: Lids}     \\ \hline\hline
$R_1$       & \multicolumn{2}{c||}{0.538} & $R_1^*$      &(cm)   & 7.1  \\
$R_2$       & \multicolumn{2}{c||}{1.538} & $R_2^*$      &(cm)   & 20.3 \\
$R_{12}$    & \multicolumn{2}{c||}{1.038} & $R_{12}^*$   &(cm)   & 13.7 \\
$H$         & \multicolumn{2}{c||}{2.114} & $H^*$        &(cm)   & 27.9 \\
$\Omega_1$  & \multicolumn{2}{c||}{3.003} & $\Omega_1^*$ & (rpm) & 200  \\
$\Omega_2$  & 0.488 & 0.400               & $\Omega_2^*$ & (rpm) & 26   \\
$\Omega_3$  & 0.488 & 1.367               & $\Omega_3^*$ & (rpm) & 26   \\
$\Omega_4$  & \multicolumn{2}{c||}{0.488} & $\Omega_4^*$ & (rpm) & 26   \\
  \end{tabular}
    }
  \end{minipage}
  \qquad\qquad
  \begin{minipage}[c]{0.4\textwidth}
\begin{center}
{\setlength{\unitlength}{1.0in}
   \begin{picture}(2.0,3.)( -0.0,-.0)
      \put(0.00,0.00){\psfig{figure=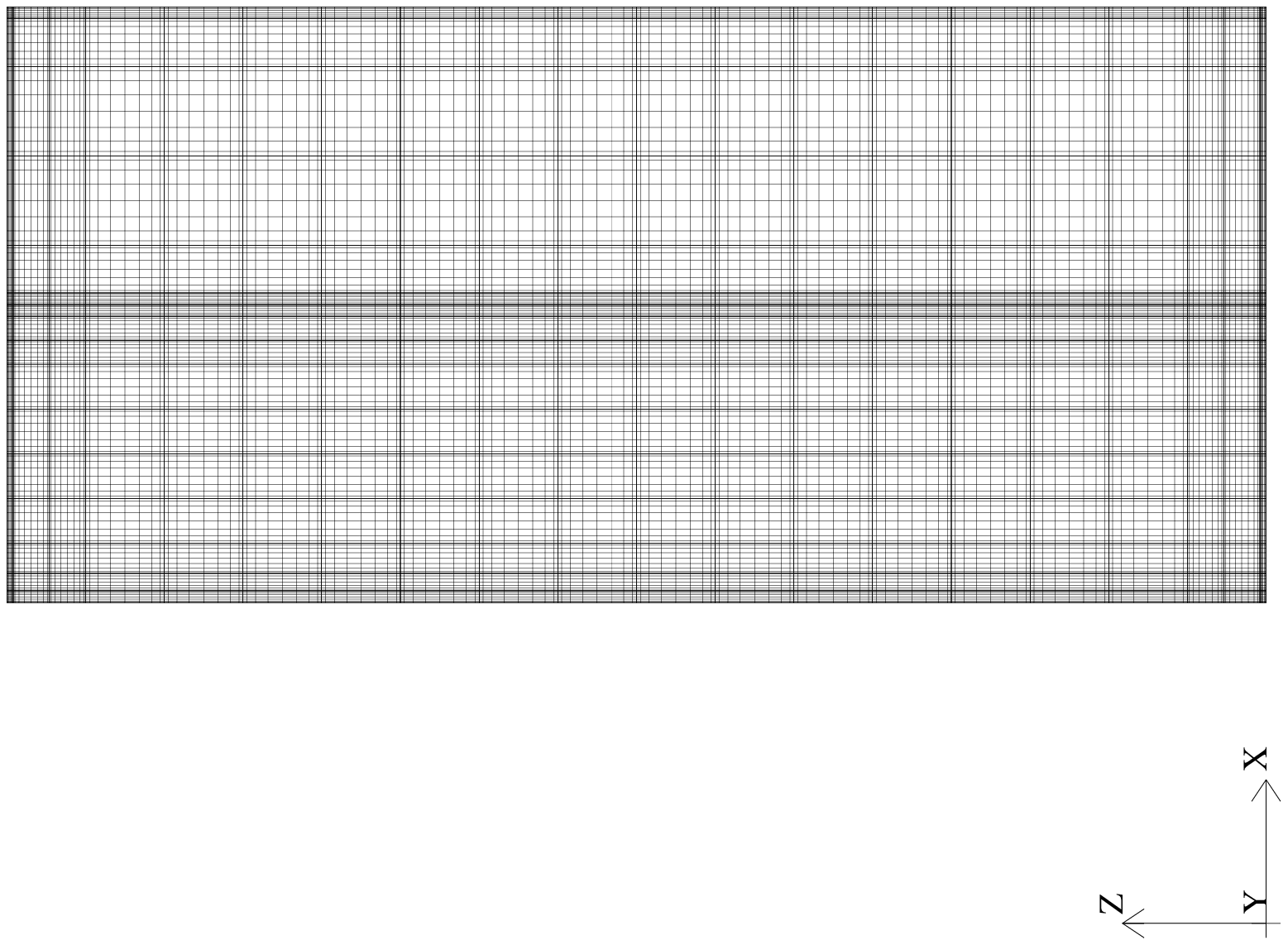,angle=-90,width=2in}}
      \put(.50,-0.15){$r=R_1$}
      \put(1.25,-0.15){$R_{12}$}
      \put(1.9,-0.15){$R_{2}$}
      \put(.25,2.65){$z=H$}
      \put(.45,1.3){$\Omega_1$}
      \put(2.1,1.3){$\Omega_2$}
      \put(0.95,2.8){$\Omega_3$}
      \put(1.6,2.8){$\Omega_4$}
\end{picture}}
\end{center}
    \vspace{-0.2in}\mbox{}
  \end{minipage}
  \caption{The geometry and rotation parameters
  for the computational cases with lids and rings at $Re=6190$ and
  experimental setup with lids at $Re=9270$
  along with the drawing of the cut of 3D computational mesh at $\theta=0$
  for the case with rings.
 {\it Note the clustering of the gridlines at the boundaries
   of the spectral elements whose location and dimensions are chosen
   to resolve efficiently boundary layers
   and `step' changes in angular velocity between cylinders and rings. }
  } \label{t:param}
\end{table}

\subsection{Numerical Technique}
\label{ss:numer}

The axisymmetric version of equations
(\ref{e:MHD:V_r}--\ref{e:MHD:div_V:axi}) and fully three-dimensional
version in cartesian coordinates has been solved numerically with
the spectral-element code Nek5000 developed and supported by Paul
Fischer and collaborators \citeaffixed[and references
within]{FOC08,FLL07}{see}.

The temporal discretization in Nek5000 is based on a semi-implicit
formulation in which the nonlinear terms are treated explicitly in
time and all remaining linear terms are treated implicitly. In
particular, we used either a combination of $k$th-order backward
difference formula (BDF$k$) for the diffusive/solenoidal terms with
extrapolation (EXT$k-1$) for the nonlinear terms or the
operator-integration factor scheme (OIFS) method where BDF$k$ is
applied to the material derivative with the explicit fourth-order
Runge-Kutta scheme being used for the resulting pure advection
initial value problem.

With either the BDF$k$/EXT$k-1$ or OIFS approach, the remaining
linear portion of time advancement amounts to solving an unsteady
Stokes problem.  This problem is first discretized spatially using
spectral-element method (SEM) and then split into independent
subproblems for the velocity and pressure in weak variational form.
The computational domain is decomposed into $K$ non-overlapping
subdomains or elements, and within each element, unknown velocity
and pressure are represented as the tensor-product cardinal Lagrange
polynomials of the order $N$ and $N-2$, correspondingly, based at
the Gauss-Lobatto-Legendre (GLL) and Gauss-Legendre (GL) points.
This velocity-pressure splitting and GLL-GL grid discretization
requires boundary condition only for velocity field and avoids an
ambiguity with the pressure boundary conditions in accordance with
continuous problem statement.

The discretized Stokes problem for the velocity update gives a
linear system which is a discrete Helmholtz operator. It comprises
the diagonal spectral element mass matrix with spectral element
Laplacian being strongly diagonally dominant for small timesteps,
and therefore, Jacobi (diagonally) preconditioned conjugate gradient
iteration is readily employed.  Then the projection of the resulting
trial viscous update on divergence-free solution space enforces the
incompressibility constraint as the discrete pressure Poisson
equation is solved by conjugate gradient iteration preconditioned by
either the two-level additive Schwarz method or hybrid
Schwarz/multigrid methods.  Note that we used
dealising/overintegration where the oversampling of polynomial order
by a factor of $3/2$ was made for the exact evaluation of quadrature
of inner products for non-linear (advective) terms.

The typical axisymmetric case with rings at high Reynolds number of
$Re=6200$ (see figure~\ref{f:lid:ring}b) required the spacial
resolution with polynomial order $N=10$ and number of spectral
elements $K=320$ (cf.~drawing for table~\ref{t:param}) and was
computed with timestep $\Delta{t}=10^{-3}$ for the duration of
$t\sim300$, while the axisymmetric run with lids at the same $Re$
(figure~\ref{f:lid:ring}a) had $N=8$, $K=476$,
$\Delta{t}=5\times10^{-3}$ and $t\sim500$.  The corresponding
three-dimensional cases with rings and lids had $N=11$, $K=9600$,
$\Delta{t}=6.25\times10^{-4}$, $t\sim280$ and $N=9$, $K=14280$,
$\Delta{t}=6.25\times10^{-4}$, $t\sim180$, respectively.  Note that
in order to facilitate time advancement and minimize CPU
requirements, the final output from another cases, e.g.~with lower
Reynolds number $Re$, was used as initial conditions for some of the
computations with higher $Re$, and the corresponding axisymmetric
cases with small random non-axisymmetric perturbation was a starting
point for most of our fully 3D computations.  Apart from CPU
savings, the usage of the perturbed axisymmetric solution obtained
in {\it cylindrical formulation}
(\ref{e:MHD:V_r}--\ref{e:MHD:div_V:axi}) as initial condition for 3D
computations at low Reynolds numbers ($Re=620$) served as an
additional validation of the code setup due to the convergence of
the fully 3D results computed in {\it cartesian formulation} back to
the unperturbed axisymmetric steady state initial condition (see
also subsection~\ref{ss:amf}).

Finally, the step change of angular velocities that mimics its
transition in the gaps or grooves between the cylinders and
horizontal boundaries as well as between the inner and outer ring in
Princeton MRI liquid gallium experiment \cite{Sc08}  was modelled
within one spectral element of the radial size $L_g=0.020$ by
ramping power law function of radius with an exponent that was
varied in the range from 4 to $N-1$ without noticeable effect on the
flow.

\section{Results}
\label{s:result}

Let us first start with examination of effects of horizontal
boundary conditions on flow pattern in general and Ekman circulation
in particular before moving to a comparison with the experiment and
examination of angular momentum transport in the cylindrical
annulus.

\subsection{Horizontal Boundary Effects}
\label{ss:BC}

Here we contrast two type of horizontal boundary conditions with an
ideal baseline case of circular Couette flow (CCF). Being zero in
the ideal case, we argue that the unbalance between `centrifugal'
rotation and centripetal pressure gradient determines the fate of
the radial flow along horizontal boundaries in the cylindrical
annulus.



In the ideal case of CCF, the sheared circular motion is balanced by
centripetal pressure gradient. To be precise, the ideal CCF is the
following exact solution of equations
(\ref{e:MHD:V_r}--\ref{e:MHD:div_V:axi}) for periodic (or
stress-free) horizontal boundary conditions:
\begin{eqnarray}    \label{e:CCF}
\qquad V_r = V_z = 0 \qquad V_\theta = r \Omega_C(r) = A \: r +
\frac{B}{r}
\nonumber\\
p_C(r) = \int^r \frac{V_\theta^2}{r} d r = \frac{A^2 r^2}{2} -
\frac{B^2}{2 r^2} + 2 A B \log r + \mbox{Const}
\end{eqnarray}
Here the constant $A$ given by equation (\ref{e:BC:OC}) is
proportional to the increase in axial angular momentum,
\begin{equation}    \label{e:L}
{\cal L}=r V_\theta = \Omega \: r^2
\end{equation}
outward between the cylinders while the constant $B$ is set by
shear-generating angular velocity drop between them. The
figure~\ref{f:CCF} shows CCF azimuthal velocity $V_\theta$ (dashed),
angular velocity $\Omega_C$ (solid), axial angular momentum $\cal L$
(dash-dotted) and negative of pressure, $-p_C$ (dotted,) for the
non-dimensional parameters given in table~\ref{t:param}.
\begin{figure}
  \centerline{\includegraphics[width=4.5in]{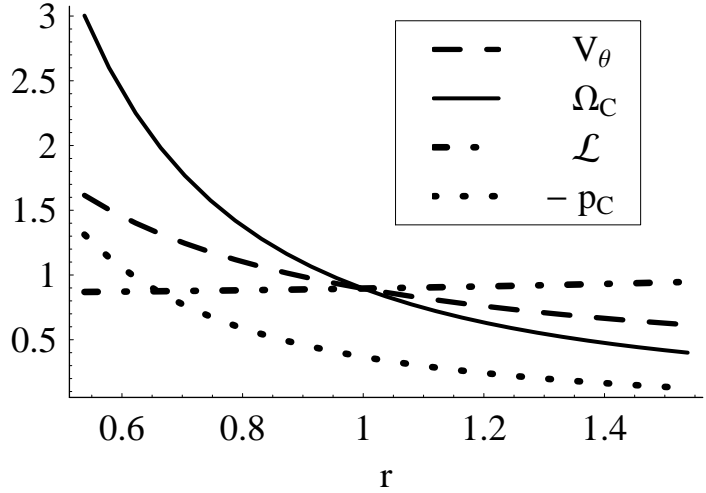}}
  \caption{The CCF azimuthal velocity $V_\theta$ (\broken),
  angular velocity $\Omega_C$ (\full), axial angular momentum ${\cal L}=r V_\theta$ (\chain) and
  minus pressure $P_C$ (\dotted) versus radius.
   {\it  Note the monotonically increasing angular momentum and decreasing
   angular velocity with radius for centrifugally stable circular Couette flow
   where the `centrifugal' rotation balances the centripetal pressure
   gradient leading to zero radial and axial velocities.} }
 \label{f:CCF}
\end{figure}
Since we are primarily interested in further MRI studies, the
baseline flow has to be centrifugally stable, i.e. with angular
momentum $\cal L$ increasing outward (for $\Omega>0$), and
therefore, satisfying Rayleigh criterion
\begin{equation}
 \dod{{\cal L}^2}{r}>0
 \label{e:Rayleigh}
\end{equation}
which is the case in this ideal CCF (dash-dotted line in
figure~\ref{f:CCF}). In order to maintain rotation with shear in
this virtual experiment with periodic horizontal boundaries, the
positive axial torque ${\cal T}_C$
\begin{equation}    \label{e:TC}
\fl {{\cal T}}_C = \int_A \left( \vec{r} \times ( d \vec{A} \bcdot
\boldsymbol{\tau} ) \right)_z = \int_0^H \int_0^{2\pi} d z d \theta
\frac{r^3}{Re} \left. \dod{}{r} \frac{V_\theta}{r}
\right|_{V_\theta=r\Omega_C} = \frac{4 \pi H R_1^2
R_2^2}{R_2^2-R_1^2} \frac{\Omega_1-\Omega_2}{Re}
\end{equation}
has to be applied to the inner cylinder while the outer cylinder is
kept from shear-free solid body rotation ($\Omega(r)=\Omega_1=A$,
$B={\cal T}=0$) by negative torque $-{\cal T}_C$.  Note that in the
above equation (\ref{e:TC}), $\boldsymbol{\tau}$ is the
non-dimensional shear stress tensor (see also \ref{s:app:amt}).

\subsubsection{Ekman Circulation with `Lids'}
\label{sss:Ekman}

In practice, the ideal CCF can never be realized in laboratory
experiments because of horizontal boundaries.  The simplest
realizable configuration is the one we refer to as `lids' when
horizontal boundaries are coupled to the outer cylinder
($\Omega_3=\Omega_4=\Omega_2$). To see how flow changes in the
presence of lids that rotate with outer cylinder, let us imagine
that these lids were inserted impulsively into fluid with ideal CCF
profile given by equation (\ref{e:CCF}) and plotted as a solid line
in figure~\ref{f:CCF} for $\Omega_1$ and $\Omega_2$ from
table~\ref{t:param}.  Keeping the most important terms in the
axisymmetric form of the equation (\ref{e:MHD:V_r}) gives
\begin{equation}
  \dod{V_r}{t} = \Omega^2 r - \dod{p}{r} + \frac{1}{Re}\dsods{V_r}{z} + \cdots
  \label{e:Lid:V_r}
\end{equation}
where we used $V_\theta=r \Omega$.  For the initial condition of CCF
(\ref{e:CCF}), the left-hand side of equation (\ref{e:Lid:V_r}) is
equal to zero everywhere outside the lids which is also consistent
with zero radial flow $V_r=0$. This zero radial flow also results in
zero diffusion term $\frac{1}{Re}\dsods{V_r}{z}$ in equation
(\ref{e:Lid:V_r}) and zero net radial force $\Omega^2 r -
\dod{p}{r}$.   The latter results from the exact CCF balance between
(positive) `centrifugal' rotation term $\Omega_C^2 \: r$ and
(negative) centripetal pressure gradient term $-\dod{p}{r}$ in
equation (\ref{e:Lid:V_r}).

Instead of initial ideal CCF angular velocity $\Omega_C$
(\ref{e:CCF}), the flow next to the lids now rotates with a smaller
angular velocity of the outer cylinder
($\Omega_2=\Omega_3=\Omega_4<\Omega_C$).  However, the centripetal
pressure gradient is still set by the bulk rotation of the rest of
the fluid and therefore, becomes suddenly larger than the
`centrifugal' rotation of fluid next to the lids,
i.e.~$\dod{p}{r}=\Omega_C^2 r>\Omega^2 r$. As a result of this
angular momentum deficit of near-wall fluid, the centripetal
pressure gradient prevails over rotation term in (\ref{e:Lid:V_r}).
Therefore, the net radial force becomes non-zero and negative,
$\Omega^2 r - \dod{p}{r}<0$, resulting in negative sign of
$\dod{V_r}{t}$ (\ref{e:Lid:V_r}) and therefore, in a formation of
the Ekman layer with an inward radial flow ($V_r<0$) in the vicinity
of the lids.

\begin{figure}
  {\centering
  \subfloat[Lids]{\includegraphics[width=3in]{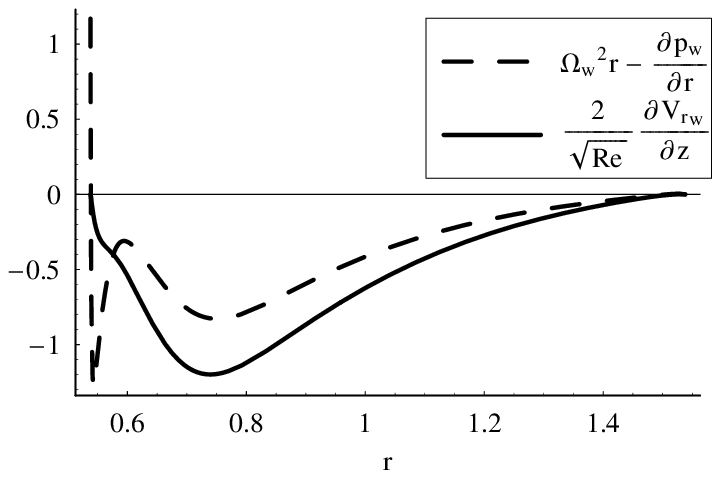}
  }
  \hspace{0.005in}
  \subfloat[Rings]{\includegraphics[width=3in]{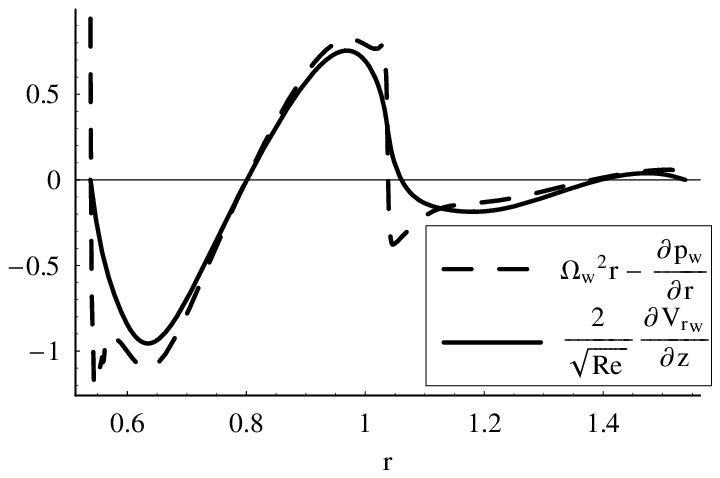}
  }
  \caption{Steady state scaled radial wall shear (\full) and
   near-wall net radial force (\broken)
   for $Re=620$ in the case of lids (a) and rings (b).
   {\it  The definite negative net radial force in the lids case (a) result
   in the inward radial Ekman flow with negative radial wall shear
   being disrupted in
   the case of rings (b) by the alternating sign of the net radial force
   that correlates well with the sign of the wall shear and thus
   with the alternating directions of Ekman flows.}} }
\label{f:dvrdt:lid:ring}
\end{figure}
The figure~2(a) 
confirms that the net radial
force near, e.g.~the lower horizontal surface $z=0$, \ $\Omega_2^2 r
- \left. \dod{p}{r} \right|_{z=0}$ (dashed) is negative, as well as
the scaled poloidal wall shear
$\frac{2}{\sqrt{Re}}\left.\dod{V_r}{z}\right|_{z=0}$. The latter
means that the $z$-derivative of $V_r$ is negative at the lower lid
which in turn results in a decrease of radial velocity with the
increase of height $z$ from noslip zero value at the lid, $V_r\Big
|_{z=0}=0$ (\ref{e:BC}) to negative values associated with the
inward Ekman flow.  Thus the deficit of angular momentum in the
near-wall fluid of the Ekman layer results in unbalanced centripetal
pressure gradient set by the bulk rotation of the rest of the flow
outside the layer and drives the Ekman flow radially inward.

\begin{figure}
  \centerline{\includegraphics[width=4.5in,angle=90]{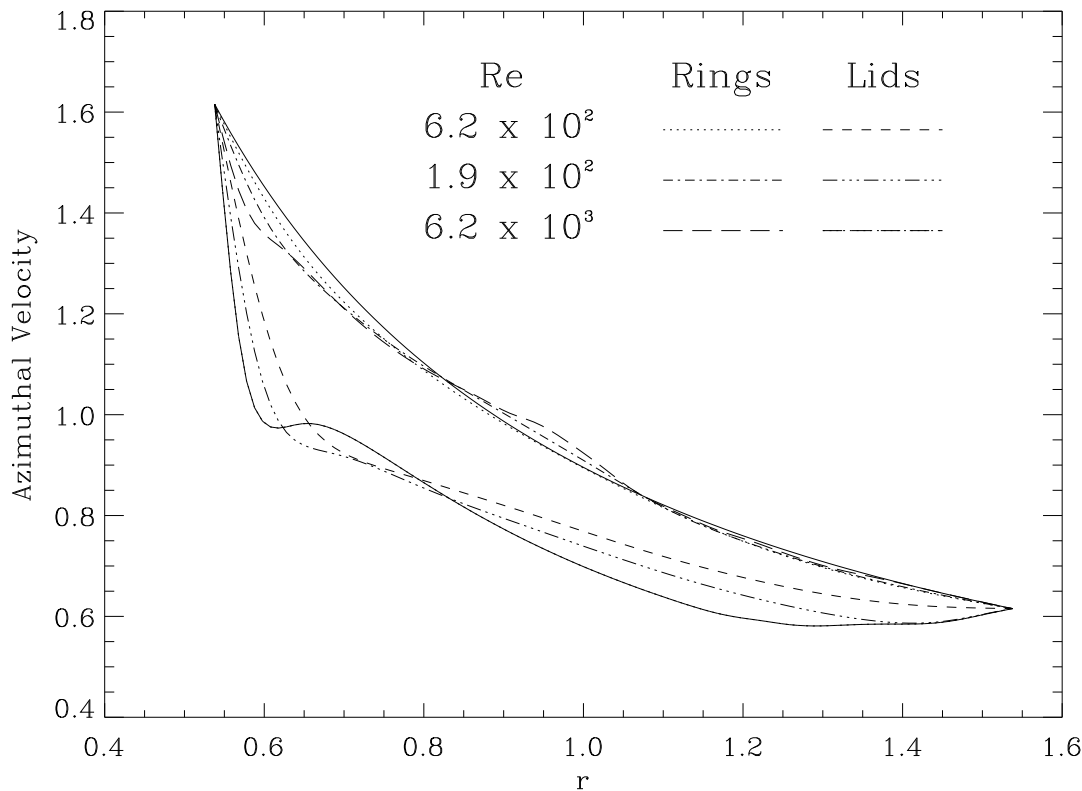}}
  \caption{Azimuthal velocity $V_\theta$ versus radius for the
  circular Couette flow (\full) and instantaneous axisymmetric
  profiles at $z=\frac{H}{4}$ in the case of lids for the
  series of Reynolds numbers $Re=620$ (\dashed), 1900
  (dash-triple-dot) and 6200 (\dashddot), and in the case of rings
  for the same Reynolds numbers: (\dotted), (\chain) and (\broken),
  respectively.
   {\it  The Ekman-circulation induced momentum deficiency in azimuthal velocity
   profiles in the cases with lids is greatly diminished
   by the particular choice of angular velocities of
   independently rotating rings.}}
 \label{f:vta}
\end{figure}
To summarize, the presence of slower rotating lids disrupt the
initial ideal CCF equilibrium between centrifugal rotation and
centripetal pressure gradient set by rotation of, respectively, lids
and bulk of the flow. This leads to the negative net radial force,
$\Omega^2 r - \dod{p}{r}<0$ and inward Ekman flow, $V_r<0$ owning to
$\dod{V_r}{t}<0$. As time grows, so does the magnitude of negative
radial velocity in the Ekman layer and, eventually, the diffusion
term $\frac{1}{Re}\dsods{V_r}{z}$ (\ref{e:Lid:V_r}) in the Ekman
boundary layer of the width $\Delta{z}\sim O(\sqrt{Re})$ becomes of
the same order (i.e.~${\sim}O(1)$)  as the net radial force that
results from two other terms in (\ref{e:Lid:V_r}). Thus the
diffusion effects finally balance the rotation momentum deficit of
the fluid in Ekman boundary layers near the lids in the saturation
steady state (see also \ref{s:app:EC}).

To check consistency of this argument, the saturation magnitude of
$\dod{p}{z}$ across the layer is verified to be more than an order
of magnitude smaller than the corresponding $\dod{p}{r}$ which
confirms that saturation centripetal pressure gradient $-\dod{p}{r}$
is indeed set by the bulk rotation of the fluid outside the Ekman
boundary layers. The saturation bulk rotation can be illustrated by
the instantaneous saturation profiles of azimuthal velocity shown in
figure~\ref{f:vta} for the steady cases with lids for $Re=620$
(dashed) and 1800 (dash-triple-dot), and unsteady case with lids of
$Re=6200$ (dash-double-dot) at $z=\frac{H}{4}$. Interesting that for
the range of Reynolds numbers considered, the effect of the increase
of Reynolds number is minor in comparison with significant azimuthal
momentum deficiency resulted from the change of horizontal boundary
conditions from initial ideal CCF (solid) to the cases of Ekman
flows over lids.

\begin{figure}
  \centering
  \subfloat[Lids]{\includegraphics[width=2.5in]{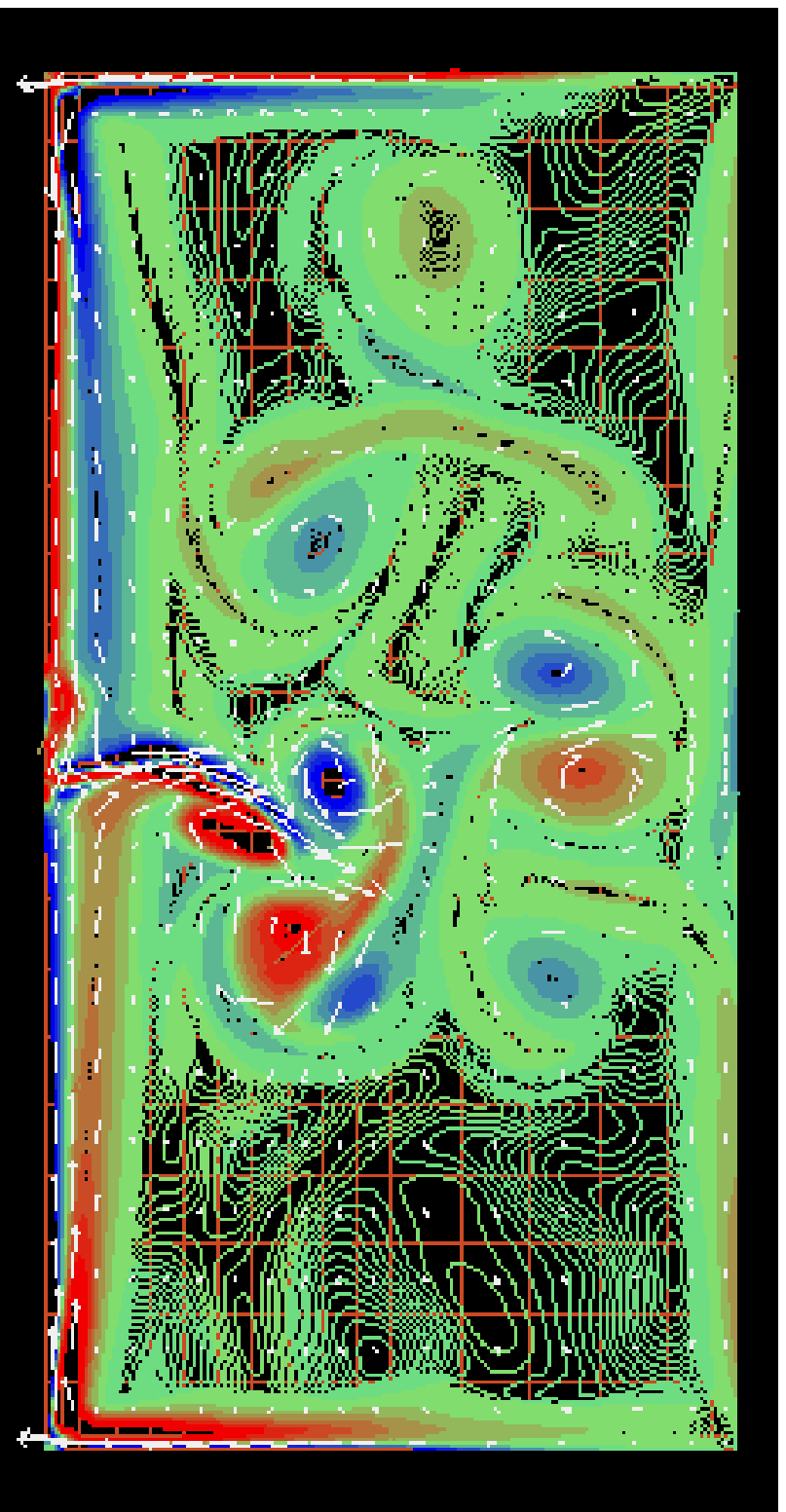}
  }
  \hspace{0.5in}
  \subfloat[Rings]{\includegraphics[width=2.5in]{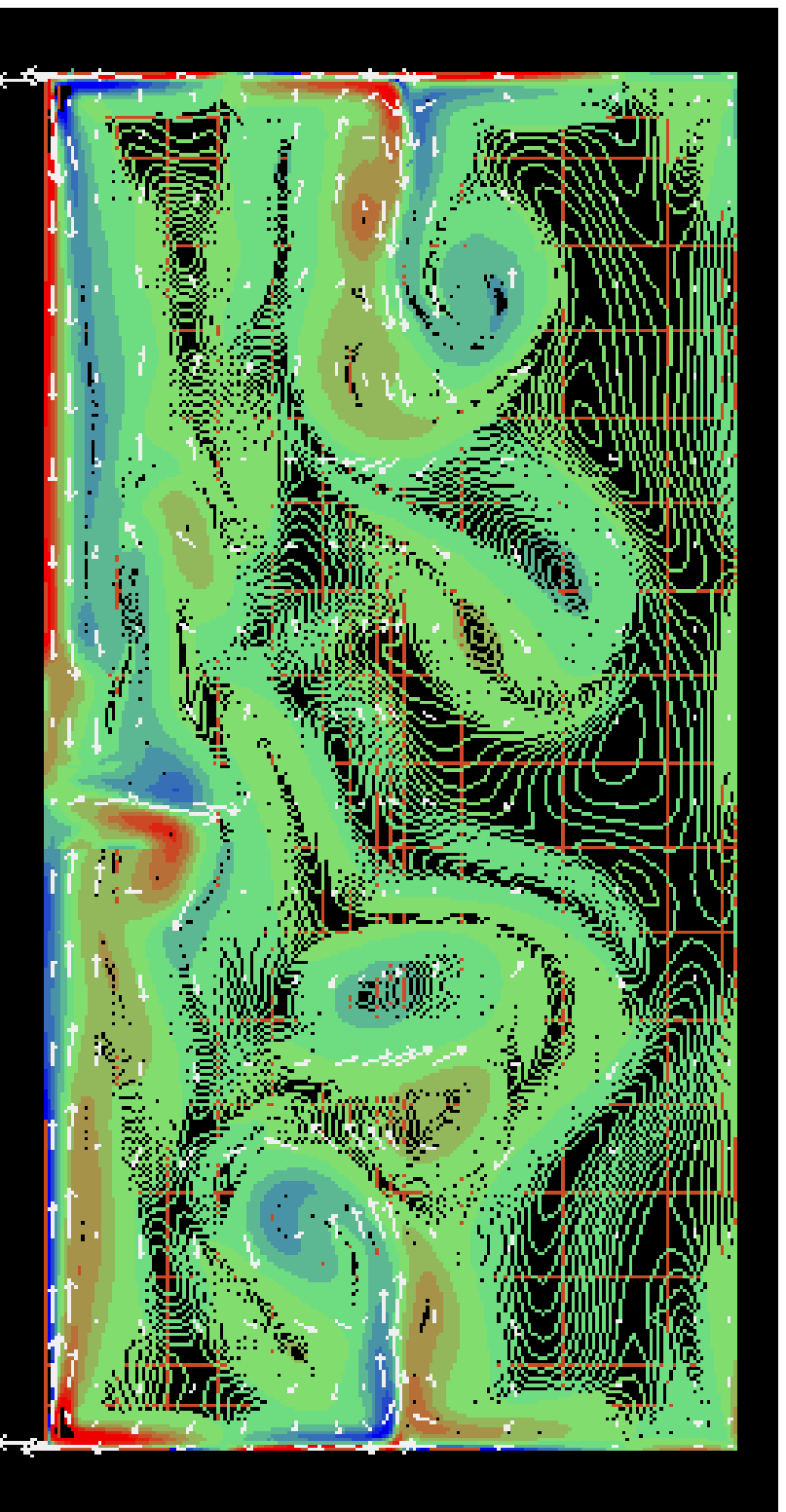}
  }
  \caption{Instantaneous contours of azimuthal vorticity and vector field of poloidal velocity
   for $Re=6200$ in the case of lids (a) and rings (b).
   {\it The Ekman circulation and outward radial jet near the midplane
   in the case of lids is severly disrupted in the setup with rings
   due to alternating inward-outward Ekman flows.}}
 \label{f:lid:ring}
\end{figure}

Owing to momentum deficiency of the near wall fluid, the higher
centripetal pressure gradient drives the inward Ekman flows along
the lids that result in EC in the cylindrical annulus.  In order to
further illustrate the phenomenon of EC due to horizontal
boundaries, we have plotted the contours of azimuthal vorticity
$\omega_\theta$ and vector plot of poloidal velocity $(V_r,V_z)$ in
figure~\ref{f:lid:ring}(a) for the case of $Re=6200$ with the former
given by
\begin{equation}
  \omega_\theta=\dod{V_r}{z}-\dod{V_z}{r}  \qquad\qquad
  \omega_\theta {\Big|_{z=0}}=\left.\dod{V_r}{z}\right|_{z=0} \quad
  \omega_\theta {\Big|_{r=R_1}}=-\left.\dod{V_z}{r}\right|_{r=R_1}
  \label{e:omega:theta}
\end{equation}
where noslip conditions (\ref{e:BC}) along the walls has been used.
Here the vorticity contours are coloured from blue
($\omega_\theta<0$) to red ($\omega_\theta>0$).  Note that this
change of colours from blue to red (through green colour whenever it
is visible) shows the locus of zero vorticity that gives approximate
location of a jet or jet-like features in the flows along the lids
at $z=0$ or $z=H$ with extremum in $V_r$ and in the flows along the
cylinders at $r=R_1$ or $r=R_2$ with minimum or maximum in $V_z$
(\ref{e:omega:theta}). In figure~\ref{f:lid:ring}(a), we observe the
Ekman boundary layers along the lids with vorticity contours
changing their colours from blue ($\omega_\theta<0$) to red at the
lower lid and from red ($\omega_\theta>0$) to blue at the upper lid.
In both instances, this change of colour shows the locus of minimum
in $V_r<0$ or location where inward Ekman flow is the strongest.
Similarly, the change of colours near the inner cylinder surface
shows the opposing vertical jet-like flows along the inner cylinder
that merge near the midplane $z=H/2$ and owing to continuity
(\ref{e:MHD:div_V:axi}), form strong outward radial  jet. At these
high Reynolds numbers, beyond $Re\sim1800$, the radial jet becomes
unsteady and starts to oscillates breaking into pairs of vortices or
to be precise, into pairs of vortex rings that move toward the lids
drawn by the mass loss in the Ekman layers and thus closing the EC
cycle.

Summing up the flow pattern in the case of lids, we conclude that
because of the deficit of rotation momentum in Ekman layers, the
fluid is pushed centripetally in these layers along the lids and
further along the inner cylinder with the subsequent formation of
the strong outward radial  jet that eventually transport fluid back
to the lids and closes the cycle of the EC (see also~\ref{s:app:EC}
and section~\ref{ss:amf}).

\subsubsection{Ekman Circulation Disruption Due to `Rings'}
\label{sss:rings}

When each horizontal boundary is split into a pair of rings that
rotate independently with the angular velocities $\Omega_3$ and
$\Omega_4$ (table~\ref{t:param}), the bulk rotation and resulting
centripetal pressure gradient is restored back to that of the CCF.
The restored profiles of azimuthal velocity in the cases with rings
are shown in figure~\ref{f:vta} for the same Reynolds numbers as for
the cases with lids, namely for the steady cases of $Re=620$
(dotted) and 1800 (dash-dot) and unsteady case of $Re=6200$ (long
dash). Along with the restoration of the bulk rotation back to that
of the CCF, we observe other major differences between the cases
with lids (a) and rings (b) in the flow field structure illustrated
in figure~\ref{f:lid:ring}.

Instead of a single outward radial jet and inward Ekman flows along
the lids, figure~\ref{f:lid:ring}(b) shows alternating
inward-outward Ekman flows along the rings that, as we describe
below, produce strong vertical jets near $r=R_{12}$ and a weaker
outward radial jet near midplane $z=H/2$. The alternating
inward-outward Ekman flows along, e.g., the lower inner and outer
rings ($z=0$), are also evident in
figure~2(b) 
where the scaled poloidal shear
$\frac{2}{\sqrt{Re}}\left.\dod{V_r}{z}\right|_{z=0}$ (solid) is
plotted as a function of radius for $Re=620$.  The radial locations
of zero shear on the inner and outer ring, ${R_s}_3$ and ${R_s}_4$,
respectively, in this case are found to be
\begin{eqnarray}
\fl \qquad {R_s}_3=0.801 \quad {R_s}_4=1.395 \qquad \mbox{such that}
\ \left.\dod{V_r}{z}\right|_{(r,z)=({R_s}_i,0)}=0 \quad \mbox{where}
\ i=3,4
  \label{e:zeros:shear}
\end{eqnarray}
We observe that the scaled poloidal shear
$\frac{2}{\sqrt{Re}}\left.\dod{V_r}{z}\right|_{z=0}$ is negative
between $r=R_1$ and $r={R_s}_3$ and between $r\approx{R_{12}}$ and
$r={R_s}_4$. Similar to the case with lids
(figure~2a), 
 this negative $z$ derivative of
$V_r$ means that $V_r<0$ and the Ekman flow along these portions of
rings is directed radially inward.  Likewise, the positive radial
velocity or outward Ekman flow between $r={R_s}_3$ and
$r\approx{R_{12}}$ and between $r={R_s}_4$ and $r=R_2$ corresponds
to positive poloidal
shear in figure~2(b). 
Furthermore, as in the case of lids
(figure~2a), 
the signs and zeros of the scaled
poloidal shear and radial velocity correlate well with that of the
net radial force $\Omega^2 r - \dod{p}{r}$
(dashed line in~figure~2b). 
In addition, these radial locations of the reversals of net radial
force and of Ekman flows near $r={R_s}_3$ and $r={R_s}_4$
(\ref{e:zeros:shear}) coincide within upto 2\% with the radial
locations $R_3$ and $R_4$ where the ideal CCF angular
velocity~(\ref{e:CCF}) matches the angular velocity of the inner and
outer ring $\Omega_3$ and $\Omega_4$ (table~\ref{t:param}), namely
\begin{eqnarray}
\fl \qquad R_3=0.793 \quad R_4=1.369 \qquad \mbox{such that} \quad
\Omega_i=\Omega_C(R_i) \quad \mbox{where} \quad i=3,4
  \label{e:R3:R4}
\end{eqnarray}

This strong correlation of reversals of the net radial force with
reversals of Ekman flow at $r={R_s}_3$ and $r={R_s}_4$
(\ref{e:zeros:shear}), coincidental with local CCF rotation at
$r=R_3\approx{R_s}_3$ and $r=R_4\approx{R_s}_4$, is completely
consistent with our argument that the balance and unbalance between
`centrifugal' rotation and centripetal pressure gradient determines
the fate of the radial flow along horizontal boundaries.  Namely,
the zero radial velocity at $r={R_s}_3$ and $r={R_s}_4$ results from
the CCF-like balance of centripetal pressure gradient $-\dod{p}{r}$
and `centrifugal' rotation $\Omega^2 r\approx\Omega_C^2 r$
(\ref{e:CCF}) since $R_3\approx{R_s}_3$ and $R_4\approx{R_s}_4$
(\ref{e:zeros:shear}--\ref{e:R3:R4}). Moreover, a monotonic decrease
of $\Omega_C$ (\ref{e:CCF}) with increase of $r$ (solid line
in~figure~\ref{f:CCF}) means that the near-wall fluid rotation at
angular velocities of the rings $\Omega_3$ and $\Omega_4$ is locally
slower (faster) than that of CCF for the radius $r$ that is smaller
(bigger) than $r\approx{R_s}_3$ and $r\approx{R_s}_4$,
correspondingly. Thus near-wall fluid rotation momentum deficit
(excess) results, respectively, in the negative (positive) sign of
the net radial force $\Omega^2 r - \dod{p}{r}$ and therefore,
negative (positive) sign of radial velocity $V_r$ in
figure~\ref{f:lid:ring}(b) and poloidal shear
$\frac{2}{\sqrt{Re}}\left.\dod{V_r}{z}\right|_{z=0}$ in
figure~2(b) 
 for the radial location $r$ that is
smaller (larger) than $r\approx{R_s}_3$ and $r\approx{R_s}_4$.


In summary,  the angular velocities of inner and outer rings
($\Omega_3$ and $\Omega_4$) set the CCF-like equilibrium radii
($r\approx{R_s}_3$ and $r\approx{R_s}_4$) by matching locally to
monotonically decreasing CCF-like profile of bulk flow rotation.
The near-wall fluid over the portions of the rings that have a
smaller radius $r$ than these CCF equilibrium radii experience
rotation momentum deficit that results in the inward Ekman flows due
to locally higher centripetal pressure gradient set by faster bulk
rotation as in the cases with lids. Conversely, when $r>{R_s}_3$ and
$r>{R_s}_4$, the bulk rotation is slower than the near-wall velocity
due to monotonic decrease of velocity profile with increase of
radius outside the Ekman layers, and the fluid has enough near-wall
rotation momentum to overcome centripetal pressure gradient and to
drive the outward Ekman flows as observed in
figure~\ref{f:lid:ring}(b).

The rest of the prominent features of the flow field in
figure~\ref{f:lid:ring}(b) like the strong vertical jets near
$r=R_{12}$ and a weak outward radial jet near the midplane $z=H/2$
are the direct consequences of these alternating inward-outward
Ekman flows along the rings. Namely, driven by rotation momentum
excess and deficit of fluid near inner and outer ring, respectively,
pairs of opposing Ekman flows along both horizontal boundaries merge
near the boundary between inner and outer ring $r=R_{12}$. Owing to
continuity (\ref{e:MHD:div_V:axi}), these pairs of colliding Ekman
flows with, presumably, equal linear radial momentum, launch the
opposing vertical jets near the ring boundary $r=R_{12}$ that become
unsteady with the increase of Reynolds number and break into vortex
pairs or vortex rings. Similarly, the Ekman flows along lower and
upper inner rings due to the rotation momentum deficit are pushed
into the corners with the inner cylinder  and further along the
inner cylinder until they merge near the midplane $z=H/2$ to form a
outward radial jet as in the case with lids. But contrary to the
cases with lids, the outward radial jet now is significantly weaker
owning to the fact that the effective Reynolds number for these
flows are smaller than in the cases with lids due to the smaller
characteristic length scale (${R_s}_3-R_1<L$) and velocity scale
($\Omega_3 {R_s}_3 - \Omega_1 R_1<U$) which leads to larger Ekman
numbers
$E=\frac{\nu}{\Delta{\Omega}\:L^2}=\frac{U/(\Delta{\Omega}\:L)}{Re}$.

Finally, we would like to make two following comments.  First,
three-dimensional effects appear to be negligible at these Reynolds
numbers with only noteworthy difference of considerably shorter
vertical jets near the ring boundary $r=R_{12}$ as compared to the
axisymmetric cases. Second, the angular velocities of rings control
the angle and direction of the jet near this ring boundary
$r=R_{12}$. In particular, when rings are coupled together and
rotate with the outer cylinder (`lids'), the jets become the inward
Ekman flows along lower and upper horizontal boundary so the the
angle with radius vector is $\pm\pi$, correspondingly. When rings
are decoupled and rotate with the angular velocities considered
above (table~\ref{t:param}), the Ekman flows collide near the ring
boundary $r=R_{12}$ and launch the opposing vertical jets, i.e. the
angle is $\pm\pi/2$. When rings are coupled to the inner cylinder,
we have checked that the resulting Ekman flows have radially outward
direction due to the excess of the near-wall angular momentum
leading to the zero angle between the jets and radius vector in
accordance with the mechanism described above. Moreover, this angle
is expected to be sensitive to the details of the flow in the
vicinity of the ring boundary such as presence of gaps between
rings, three-dimensionality, etc.~but it is likely to be adjusted
with an appropriate choice of angular velocities of rings shifting
the equilibrium points of local CCF balance and thus regulating the
radial extent and radial linear momentum of the Ekman flows
\citeaffixed{Sc08}{cf.}. In other words, the angular velocities of
rings control EC through the net radial momentum after the collision
of Ekman flows that sets the angle at which the jets are launched
near the ring boundary $r=R_{12}$.


%
%
%

\subsubsection{Summary on Horizontal Boundary Effects}
\label{sss:BC}

The CCF-like equilibrium between `centrifugal' rotation and
centripetal pressure gradient in cylindrical annulus is impossible
to achieve experimentally due to the presence of the noslip
horizontal boundaries.  The rotation of these boundaries with either
faster inner cylinder or slower outer cylinder creates the Ekman
boundary layers with either angular momentum excess or deficit,
correspondingly, and results into either outward or inward Ekman
flows, respectively, that drive EC in the annulus.  The splitting of
the horizontal boundaries into independently rotating rings sets the
CCF-like equilibrium points by matching locally to the CCF angular
velocity, and resulting angular momentum deficit or excess leads to
the, correspondingly, inward or outward Ekman flows along the
portions of the rings with radius, respectively, smaller or larger
than the radius of these equilibrium points. The opposing Ekman
flows along the rings collide near ring boundaries and launch the
vertical jets at an angle presumably determined by the mismatch of
their linear radial momentum. This angle is expected to be sensitive
to the details of the flow structure inside and immediately near the
gaps between rings, the vertical alignment of the horizontal
surfaces of the rings, etc. and can be adjusted
by changing the angular velocities of the rings \citeaffixed{Sc08}{cf.}

\subsection{Comparison with Experiment}
\label{ss:exp}

\begin{figure}
  \centerline{\includegraphics[width=5in]{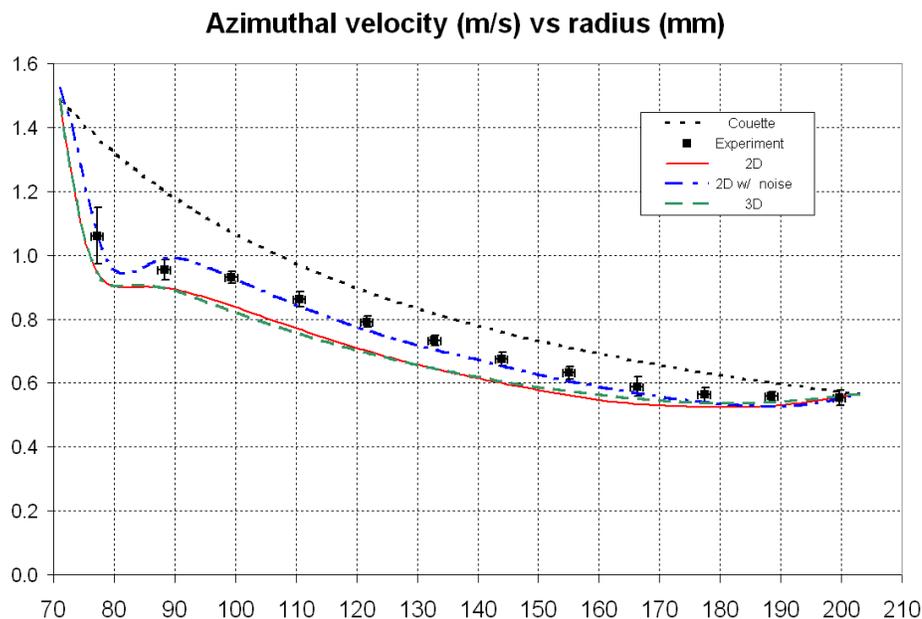}}
  \caption{Dimensional azimuthal velocity profile versus radius at $z=\frac{H}{4}$ for
  experimental data (\fullsquare) at $Re\approx{9300}$ \cite{Sc08},
  circular Couette flow (\dotted), and our numerical simulations at
  $Re=6200$: three-dimensional (\full), axisymmetric (\dashed)
  and noisy axisymmetric (\chain) cases.
   {\it Three-dimensional effects are negligible compared to axisymmetric case
   both being slightly lower than experimental profile, and the best fit
   is achieved in the axisymmetric case with random noise perturbations
   applied to the surface of inner cylinder.} }
 \label{f:vta:exp}
\end{figure}

We have collaborated with Princeton MRI liquid gallium experiment
group and conducted a comparison of our computations with their
experimental results.  Figure~\ref{f:vta:exp} shows the comparison
of our numerical results for time-averaged azimuthal velocity in the
case with lids at Re=6200 with ideal CCF profile (dotted) and
experimental measurements (squares) conducted by Schartman (2008) at
$Re\approx{9300}$.  The solid line corresponds to the axisymmetric
computation while the fully three-dimensional results are shown with
the dashed line. We observe that at this Reynolds number
($Re\approx6200$) three-dimensional time-averaged azimuthal velocity
is very close to the axisymmetric one both being upto $15\%$ lower
than the experimental data.  The difference in Reynolds number is
expected to play only a minor role in this discrepancy.

The best fit of our (axisymmetric) computations (dash-dot line) with
experimental data was realized  when the boundary conditions
(\ref{e:BC}) were perturbed with uniform random noise. The amplitude
of the noise was 5\% relative to the corresponding maximums of
axisymmetric solution without noise. The noise perturbation was
applied for the part of the computational domain boundary of one
spectral element long such that $R_1\le{r}<R_1+L_g$ where
$L_g=0.0200$ which is less than twice of the non-dimensional width
of the gap  between the inner cylinder and inner ring in the
Princeton experiment equal to 0.0114 or 1.5 mm (Schartman 2008).

The rational behind the noise perturbation of the boundary
conditions was an attempt to model an effect of the centrifugally
unstable flow in the gap between inner cylinder and inner ring. By
an accident, boundary conditions on the inner cylinder surface
($r=R_1$) were also perturbed in this computation which turned out
to be the best fit with experimental data.  The effect of this
perturbation of the inner cylinder surface boundary condition may be
similar to a random blowing. This leads us to believe that a
combination of the effects due to a run-out of inner cylinder and
due to the centrifugally unstable flow in the gaps between the inner
cylinder and inner ring may explain the discrepancy between the
simulation and experiment (see also Schartman 2008).

The further work on comparison of the simulation with experiment is
ongoing, and addition effort is needed to sort out the effects of
run-out, centrifugally unstable flow in the gaps between cylinders
and rings, vertical misalignment of horizontal surfaces of the
rings, etc.


\subsection{Torque and Angular Momentum Transport} \label{ss:amf}



Also we have studied carefully the torque behaviour and associated
angular momentum transport in the hydrodynamical setup of Princeton
MRI liquid gallium experiment (Schartman 2008) as a baseline cases
for our study of magneto-rotational instability (MRI) and MRI-driven
turbulence. In order for the MRI experimental results to have a
clear interpretation, the negative effects of the EC has to be
minimized, and the torque amplification over the CCF torque
${\cal{T}}_C$ (\ref{e:TC}) with the increase of magnetic field can
be linked directly to MRI enhancement of angular momentum transport
(AMT). Therefore, the understanding of the torque behaviour and AMT
in the baseline cases of hydrodynamical flow can not be overstated.

An application of torque to the inner cylinder results in a flow
that transports the angular momentum outward and attempts to reach a
shear-free solid body rotation with a constant angular velocity.  If
the angular velocities of other boundaries are different form that
of the inner cylinder, the resulting shear has to be maintained by
application of torques to the boundaries in order to keep the
rotation rates steady.  In the context of MRI study, the primary
interest is in the transport of the angular momentum from the inner
to outer cylinder in centrifugally stable regime and, as shown
below, in the minimization of EC and hence, in the reduction of the
net contribution of torques exerted on the horizontal boundaries.
Since the sum of all torques applied to the boundaries reflects the
time increase of interior angular momentum, this contribution from
the horizontal boundaries is equal to the sum of torques applied to
the cylinders in a steady state and has to be minimized for the
successful MRI experiment. Note that in a case of unsteady flow, the
time averaged torques are used when the statistically steady state
is reached.

\begin{figure}
  \centerline{\includegraphics[width=5.5in]{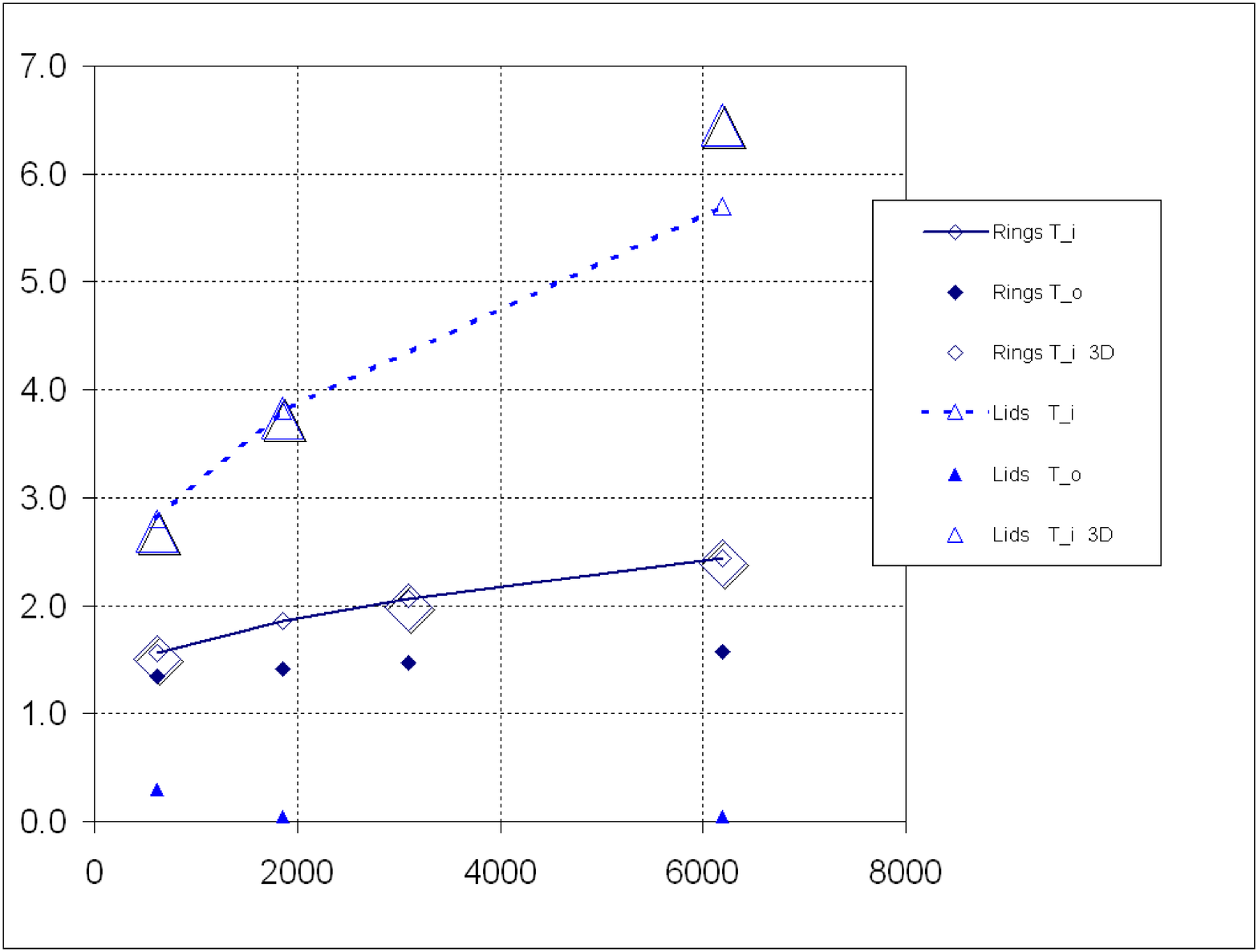}}
  \caption{The magnitudes of normalized torques applied to inner cylinder
  ($T_i= {\cal T}_1/{\cal T}_C$, open symbols) and outer cylinder
  ($T_o=-{\cal T}_2/{\cal T}_C$, filled symbols)
  versus Reynolds number for the cases with rings
  (\ \ \opendiamond \hspace{-4ex} \full \ or \fulldiamond) and with
  lids (\ \ \opentriangle \hspace{-4ex} \dotted \ or \fulltriangle)
  in axisymmetric cases while the three-dimensional data
  are plotted with large symbols.
   {\it  Being the sum of magnitudes of torques applied to the
   inner and outer cylinder,  the net torque exerted on the
   horizontal boundaries in the case of rings is significantly less
   than that in the case of lids making the former closer
   approximation to the CCF for which the net torque on horizontal
   boundaries is zero.} }
  \label{f:torque}
\end{figure}

The figure~\ref{f:torque} shows the Reynolds number dependence of
magnitudes of steady/time-averaged torque  relative to the ideal CCF
torque ${\cal T}_C$ (\ref{e:TC}) that has to be applied on inner and
outer cylinders, $T_i=\frac{{\cal T}_1}{{\cal T}_C}$ (open symbols)
and $T_o=\frac{-{\cal T}_2}{{\cal T}_C}$ (filled symbols),
respectively, in order to keep constant boundary angular velocities
(table~\ref{t:param}). Being equal to unity for the case with
periodic boundary conditions (CCF), the torque magnitudes are shown
for the cases with rings (open diamonds with solid lines and filled
diamonds) and lids (open triangles with dotted lines and filled
triangles).  All results are obtained in axisymmetric computations
except for the data plotted with large open symbols that shows the
results of fully 3D computations.  Note that being zero in the ideal
CCF, the difference between the torques applied to the inner and
outer cylinder shown by open and closed symbols, respectively,
corresponds to the sum of torques exerted on the fluid next to the
horizontal boundaries
$$
T_i-T_o=({\cal T}_1+{\cal T}_2)/{\cal T}_C
$$
due to zero net torque in steady/statistically steady state.
Evidently, the setup with rings has an advantage of a smaller
contribution to the net torque from the horizontal boundaries and of
a smaller difference between inner and outer cylinder torque
magnitudes over the setup with lids where EC is undisturbed.  We
also observe that in the range of Reynolds numbers considered the
flow makes a transition from steady axisymmetric solution at
$Re=620$ to the unsteady one at $Re=6200$ with small
three-dimensional effects. Being more significant in the case with
lids, three-dimensionality is expected to play an increasing role
with the further increase of Reynolds number.

\begin{figure}
  \centering
  \subfloat[Lids]{\includegraphics[width=2.00in]{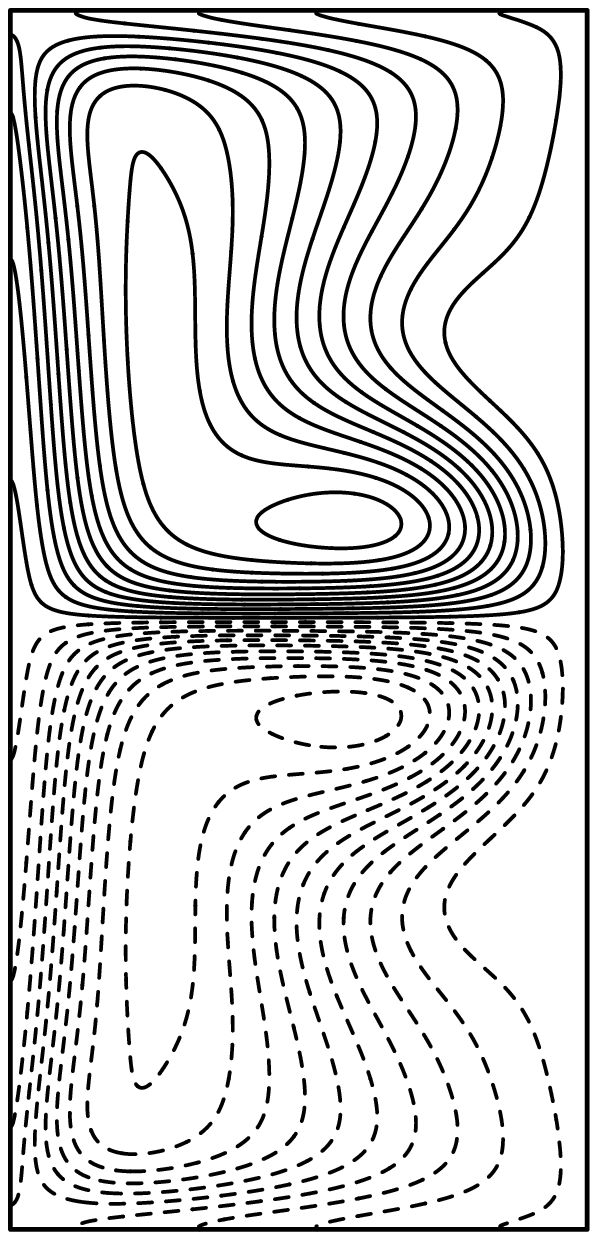}
  }
  \hspace{0.5in}
  \subfloat[Rings]{\includegraphics[width=2.00in]{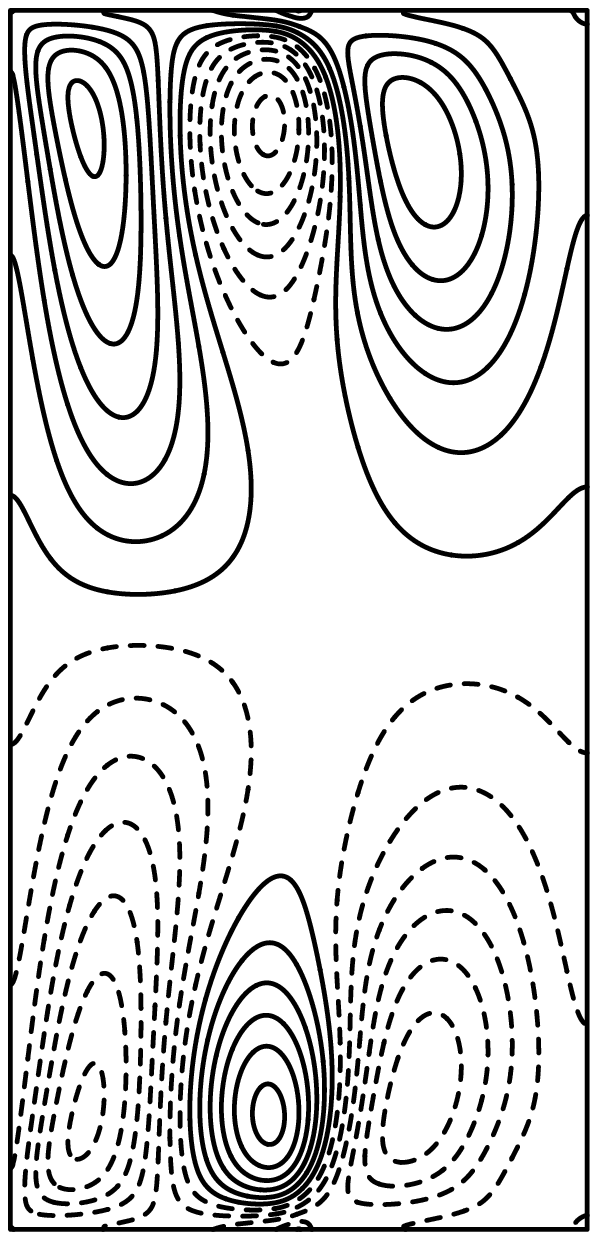}
  }
  \caption{Steady state contour lines of effective angular momentum
   flux function $\tilde{\Psi}$
   for the case of $Re=620$ with lids in the range from -3.30 to 3.30 in the increment of
   0.31 (a) and with rings in the range from -1.35 to 1.35 in the increment
   of 0.21 (b).
   {\it In the case of lids, most of the flux lines
   that originate from inner cylinder terminate at horizontal boundaries as
   opposed to the case of rings where they end up mostly at the outer cylinder
   which is similar to the CCF angular momentum transport between the
   cylinders.}}
 \label{f:amf:lid:ring}
\end{figure}

In order to illustrate a spacial variations of AMT, we have computed
an effective angular momentum flux function defined in
\ref{s:app:amt} by analogy with a streamfunction. The contours of
the effective flux function shows the (flux) lines along which the
angular momentum is transported, and the difference between the
values of the flux function at two points gives the total flux
across the segment of conical or cylindrical surfaces on which these
points lie. Figure~\ref{f:amf:lid:ring} shows steady state contour
lines of constant increment for effective angular momentum flux
function $\tilde{\Psi}$ (\ref{e:Psi:tilde}) for the case of $Re=620$
with lids (a) and rings (b).  For comparison, we note that the flux
lines of (purely viscous) AMT for the ideal CCF (\ref{e:CCF}) are
the straight lines from inner to outer cylinder along $z=$const.
Despite the fact that in both cases only a single line of
$\tilde{\Psi}=0$ (i.e.~line of symmetry) is the same as in CCF case,
the case with rings exhibits the similar transport of angular
momentum along the flux lines that mostly originate at the inner
cylinder and terminate at the outer cylinder in contract to the
termination of the flux lines at the lids.  The latter indicates
that in the cases with lids, the angular momentum transport is
mostly between the inner cylinder and the horizontal boundaries
contrary to more desirable CCF-like transport between the cylinders
observed in the cases with rings.  Also note that the similarity
between the shape of the flux lines away from the boundaries in
figure~\ref{f:amf:lid:ring} and the shape of vorticity contour lines
and poloidal vector lines in figure~\ref{f:lid:ring} can be
explained through creation of strong $V_r$ and $V_z$ components of
velocity due to Ekman flows which affects AMT flux through advective
contributions $F^a_{rz}$ and $F^v_{zz}$, respectively, given by
relations (\ref{e:amf:a:r}) and (\ref{e:amf:a:z}).

In summary, if the ultimate objective is to achieve the flow with
AMT as close to the ideal CCF as possible, the design with rings
seems to have an advantage over the setup with lids.

\section{Conclusion and Future Work}
\label{s:concl}

In this paper we have presented axisymmetric and fully
three-dimensional Navier-Stokes calculations of circular Couette
flow (CCF) in a cylindrical annulus as the first step in our study
of magneto-rotational instability (MRI) and MRI-driven turbulence.
Inspired by Princeton MRI liquid gallium experiment, we have
computed the flow field in their experimental setup for realistic
horizontal boundary conditions of `lids' and `rings' with the
increase of Reynolds number through the onset of unsteadiness and
three-dimensionality. The presented analysis of the flow field and
angular momentum transport (AMT) allowed us to propose an
explanation of the mechanism that determines the fate of the
boundary flows and Ekman circulation (EC) as a result of a
competition between the effects of `centrifugal' rotation and
pressure gradient set by rotation of, respectively, horizontal
surfaces and bulk of the flow.  In particular, with the appropriate
choice of rotation rates of the horizontal rings that control an
angle at which the vertical jets are launched near the ring
boundaries, EC can be greatly diminished and CCF-like flow can be
restored being more appropriate for the further experimental studies
of MRI saturation and enhanced AMT.  In addition, our numerical
results compare favourably with the experimental data with the
maximum deviation below 15\% being considerably smaller in the cases
with `noisy' boundary conditions.  The future work, therefore,
should involve higher Reynolds number computations with even more
detailed modelling of experimental geometry that includes among
others the effects of run-out of the inner cylinder, finite gaps
between the cylinders and rings, and vertical misalignment of
horizontal surfaces of the rings.

\ack

We acknowledge the support of National Science Foundation sponsored
Physics Frontier Centre for Magnetic Self-Organization in Laboratory
and Astrophysical Plasma (CMSO), and the use of computational
resources of Argonne Leadership Computing Facility (ALCF) operated
by Argonne National Laboratory and of the National Energy Research
Scientific Computing Center (NERSC) at Lawrence Berkeley National
Laboratory supported by the Office of Science of U.S. Department Of
Energy (DOE) under Contract No.~DE-AC02-05CH11231.  The work was
also partially supported by NASA, grant number NNG04GD90G, and by
the Office of Science of the U.S. DOE under Contract
No.~W-31-109-Eng-38. We are also grateful to Ethan Schartman,
Michael Burin, Jeremy Goodman and Hantao Ji and to Leonid Malyshkin.
Many thanks to Aspen Center for Physics and to International Centre
for Theoretical Physics, Trieste, Italy and especially to Snezhana
Abarzhi for the invitation to participate in the First International
Conference ``Turbulent Mixing and Beyond,'' encouragement and
discussions concerning this work.

\section*{References}


\bibliography{obabko}

\appendix

\section{Traditional Explanation of Ekman Circulation}
\label{s:app:EC}

Here we would like to comment that the traditional explanation of
Ekman flows and circulation involves a balance between Coriolis and
viscous forces in Ekman layers along rotating stressed boundary as
viewed from a uniformly rotating reference frame
\citeaffixed{Ba67,Gr68}{see }.  As we show below, this balance of
Coriolis and viscous forces holds, for instance, in the case of a
bulk flow outside the Ekman layers that is close to the state of
solid body rotation at, say, angular velocity of $\Omega_2$, when
the centripetal pressure gradient and centrifugal forces cancel each
other:
\begin{eqnarray}
  p \sim \frac{1}{2}\Omega_2^2 r^2 \qquad\qquad  r \Omega_2^2 +
  \dod{p}{r}\sim0
  \label{e:fict:p}
\end{eqnarray}
\citeaffixed{Gr68}{to the order of Ekman number
$E=\frac{\nu}{\Delta{\Omega}\:L^2}\sim\frac{1}{Re}$, see} and
Coriolis force is solely responsibly for the Ekman flow. In other
flows like those considered in this paper, a contribution of the
`centrifugal' rotation to the Ekman flow does not reduce solely to
Coriolis force but also includes the effects of  centrifugal forces.
Indeed, for the flow description in the non-inertial reference frame
that rotates with constant angular velocity of the noslip boundary
of the `lids' $\mathbf{\Omega_2}=\Omega_2\:\mathbf{e_z}$, the
centrifugal and Coriolis body forces has to be added to the
right-hand side of equations (\ref{e:MHD:V_r}--\ref{e:MHD:V_z}),
\begin{eqnarray}
  \boldsymbol{f} & = &
  - \boldsymbol{\Omega_2}\times(\boldsymbol{\Omega_2}\times\boldsymbol{r}) -
  2 \: \boldsymbol{\Omega_2} \times \boldsymbol{u}
  \nonumber\\
  f_r & = & \Omega_2^2 r  +  2 \: \Omega_2 \: u_\Theta
  \label{e:fict}
\end{eqnarray}
along with the replacement of the inertial frame velocity
$\mathbf{V}$ with the velocity in the non-inertial rotating
reference frame $\mathbf{u}$ leaing to the following radial momentum
equation in, e.g.,~case of axisymmetry ($\dod{}{\theta}=0$)
\begin{eqnarray}
  \dod{u_r}{t} + \boldsymbol{u} \bcdot \boldsymbol{\nabla} -
  \frac{u_\Theta^2}{r} = \Omega_2^2 r + 2 \: \Omega_2 \: u_\Theta
  + \frac{1}{Re} \left[ \triangle_{(r,z)} {u_r} - \frac{u_r}{r^2}
     \right] - \dod{p}{r}
  \nonumber
\end{eqnarray}
or
\begin{eqnarray}
  \dod{u_r}{t} = \Omega_2^2 r + \left( 2 \: \Omega_2 \: u_\Theta +
  \frac{u_\Theta^2}{r} \right) - \dod{p}{r} + \frac{1}{Re}\dsods{u_r}{r} + \cdots
  \label{e:fict:u_r}
\end{eqnarray}
Let us now compare this relation with equation (\ref{e:Lid:V_r})
where we expand the `centrifugal' term $\Omega^2r$ in powers of
azimuthal velocity deviation from the state of solid body rotation
of the lids, $U_\theta=V-r\Omega_2$:
\begin{eqnarray}
  \Omega^2 r=\frac{V_\theta^2}{r} = \frac{\left(r\Omega_2 + U_\theta\right)^2}{r}
  = \Omega_2^2 r + \left( 2 \: \Omega_2 \: U_\theta + \frac{U_\theta^2}{r} \right)
  \label{e:V_t:U_t}
\end{eqnarray}
Therefore, equation of radial momentum balance (\ref{e:Lid:V_r})
rewritten with the expansion~(\ref{e:V_t:U_t}) coincides exactly
with equation (\ref{e:fict:u_r}) in the case of axisymmetric flow
when the difference between $V_r$ and $u_r$ and between $U_\theta$
and $u_\Theta$ disappears due to the relationship between velocities
in the inertial and non-inertial reference frames:
\begin{eqnarray}
  \mathbf{V}(r,\theta,z,t) & = & r \: \Omega_2 \: \mathbf{e_\Theta}
  + \mathbf{u}(r,\Theta=\theta+\Omega_2 t,z,t)
  \nonumber\\
\fl \mbox{or} & & \nonumber \\
  \mathbf{V}(r,z,t) & = & r \: \Omega_2 \: \mathbf{e_\theta}
  \:\:\:\!\! + \mathbf{u}(r,z,t)
  \label{e:V:u}
\end{eqnarray}
In a particular case of flows that are close to solid body rotation
when considered in this description (\ref{e:fict:u_r}) equivalent to
our earlier framework (\ref{e:Lid:V_r}), the contribution of the
centrifugal and pressure gradient forces cancels out
(\ref{e:fict:p}) resulting in the steady state balance of Coriolis
and viscous forces,
\begin{eqnarray}
  0 = 2 \: \Omega_2 \: u_\Theta + \frac{1}{Re}\dsods{u_r}{r} + \cdots
  \label{e:fict:u_r:SB}
\end{eqnarray}
where we omitted higher order terms including
$\frac{u_\Theta^2}{r}\ll 2 \: \Omega_2 \: u_\Theta$ away from the
axis of rotation $r=0$ due to vanishing $u_\Theta$ at the noslip
boundary.  In more general case, the additional contributions of the
centrifugal forces $\Omega_2^2r$ and $\frac{u_\theta^2}{r}$ to the
Ekman flows has to be considered being a part of the `centrifugal'
term $\Omega^2r$ (\ref{e:Lid:V_r}, \ref{e:fict:u_r}) that is
balanced by centripetal pressure gradient and viscous forces in the
Ekman flows along the lids that drive EC in the cylindrical annulus.

\section{Angular Momentum Flux and Flux Function}
\label{s:app:amt}

Let us take the axisymmetric version of azimuthal momentum equation
(\ref{e:MHD:V_t}) with viscosity $\nu$ instead of $\frac{1}{Re}$ and
rewrite it in terms of conservation of axial angular momentum ${\cal
L}=r V_\theta$. The summation of the axisymmetric versions of
equations (\ref{e:MHD:V_t}) and (\ref{e:MHD:div_V:axi}) multiplied
by factors $-r$ and $r V_\theta$, correspondingly, gives

\begin{eqnarray}
\fl -\dod{}{t} \left( r V_\theta \right) & = & - r \left[ - V_r
\dod{V_\theta}{r} - V_z \dod{V_\theta}{z} - \frac{V_r V_\theta}{r} +
\nu \: \left(\dsods{V_\theta}{r} + \frac{1}{r} \dod{V_\theta}{r} +
\dsods{V_\theta}{z} - \frac{V_\theta}{r^2}
\right)  \right] \nonumber\\
\fl & + & r V_\theta \left[ \dod{V_r}{r} + \dod{V_z}{z} +
\frac{V_r}{r} \right]
\nonumber\\
\fl & = & \left[ r V_r \dod{V_\theta}{r} + r V_z \dod{V_\theta}{z} +
V_r V_\theta - \nu \: \left\{ r \dsods{V_\theta}{r} \right\} -\nu \:
r \left( \frac{1}{r} \dod{V_\theta}{r} - \frac{V_\theta}{r^2}
\right) -\nu \: r \dsods{V_\theta}{z} \right]
\nonumber\\
\fl & + & \left[ r V_\theta \dod{V_r}{r} + r V_\theta \dod{V_z}{z} +
V_r V_\theta \right]
\nonumber\\
\fl & = & \dod{ \left[ r V_r V_\theta \right] }{r} + \dod{ \left[ r
V_\theta V_z \right] }{z} + V_r V_\theta
- \nu \left\{ \dod{}{r} \left[ r \dod{V_\theta}{r} \right] -
\dod{V_\theta}{r} \right\}
- \nu \: r \dod{}{r} \left(
\frac{V_\theta}{r} \right) \nonumber\\
\fl & + & \dod{} {z}\left[ - \nu r \dod{V_\theta}{z} \right]
\nonumber\\
\fl & = & \dod{}{z} \left[ r V_\theta V_z - \nu r \dod{V_\theta}{z}
\right]  + \frac{\left[ r V_r V_\theta -\nu r^2 \dod{}{r} \left(
\frac{V_\theta}{r} \right) \right]}{r} + \dod{ }{r} \left[r V_r
V_\theta\right] \nonumber\\
\fl & - & \nu \left\{ \dod{}{r} \left[ \left( r^2 \dod{}{r} \left(
\frac{V_\theta}{r} \right) + V_\theta \right)  - V_\theta \right]
\right\}
\nonumber\\
\fl & = & \dod{}{z} \left[ r V_\theta V_z - \nu r \dod{V_\theta}{z}
\right]  + \frac{\left[r V_r V_\theta -\nu r^2 \dod{}{r} \left(
\frac{V_\theta}{r} \right) \right]}{r} + \dod{ }{r} \left[ r V_r
V_\theta -\nu r^2 \dod{}{r} \left( \frac{V_\theta}{r} \right)
\right]
\nonumber\\
\fl & = & \frac{1}{r} \dod{ }{r} \left( r \left[ r V_r V_\theta -\nu
r^2 \dod{}{r} \left( \frac{V_\theta}{r} \right) \right] \right) +
\dod{}{z} \left( r V_\theta V_z - \nu r \dod{V_\theta}{z} \right)
\label{e:L:amf:derivation}
\end{eqnarray}

The above equation (\ref{e:L:amf:derivation}) reflects the
conservation of axial or $z$-component of angular momentum and can
be condensed to
\begin{eqnarray}
\dod{}{t} \left( r V_\theta \right) +
 \frac{1}{r} \dod{}{r} ( r F_{rz} ) + \dod{F_{zz}}{z} & = & 0
 \label{e:amf:amte}
\\
\fl \mbox{or} \nonumber
\\
\left(  \dod{}{t} ( \bi{r} \times \bi{V} ) + \nabla_{(r,z)} \bcdot
\mathbf{F} \right)_z & = & 0
 \label{e:amf:amte:vec}
\end{eqnarray}
where total angular momentum flux tensor $\mathbf{F}$ and its
components $F_{rz}$ and $F_{zz}$ are given for axisymmetric flows by
\begin{eqnarray}
\mathbf{F} & = & \mathbf{F^a} + \mathbf{F^v} 
 \label{e:amf:total}\\
F_{rz}^a & = & r V_r V_\theta
 \label{e:amf:a:r}\\
F_{zz}^a & = & r V_\theta V_z
 \label{e:amf:a:z}\\
F_{rz}^v & = & -\nu r^2 \dod{}{r} \left( \frac{V_\theta}{r} \right)
= - r \tau_{r\theta}
 \label{e:amf:v:r}\\
F_{zz}^v & = & - \nu r \dod{V_\theta}{z} = - r \tau_{z\theta}
 \label{e:amf:v:z}
\end{eqnarray}
where superscript $a$ and $v$ denote an advective and viscous
contributions to the total flux of angular momentum, and
$\tau_{r\theta}$ and $\tau_{z\theta}$ are components of the stress
tensor $\boldsymbol{\tau}$.

By analogy with poloidal streamfunction $\psi$ that satisfies
axisymmetric version of continuity equation (\ref{e:MHD:div_V:axi})
due to, e.g., a definition
\begin{eqnarray}
V_r = \dod{\psi}{z} \qquad\qquad V_z = -\frac{1}{r} \dod{ ( r \psi
)}{r}
 \label{e:psi}
\end{eqnarray}
an angular momentum flux function $\Psi(r,z)$ for steady flows can
be introduced to satisfy the steady version of equation
(\ref{e:amf:amte}):
\begin{eqnarray}
F_{rz} = \dod{\Psi}{z} \qquad\quad\ F_{zz} = -\frac{1}{r} \dod{ ( r
\Psi )}{r}
 \label{e:Psi}
\end{eqnarray}
As the lines of constant value of poloidal streamfunction $\psi$ are
lines of constant poloidal flow rate with poloidal velocity vector
being tangent to these streamlines, the lines of constant total
angular momentum flux function $\Psi$ are the flux lines along which
angular momentum is transported, and the difference between these
values at any two points gives the total flux of angular momentum
transferred across the line that joints these points.  Note that in
the case of unsteady flow, the time-averaged quantities can be used
in statistically steady state instead of instantaneous ones. Also
note that similar to the streamfunction, the flux function $\Psi$ as
a solution of equations (\ref{e:Psi}) is defined up to a constant.
We have used the point of half height on inner cylinder as a zero
value for the flux function
\begin{eqnarray}
\Psi\left(r=R_1,z=\frac{H}{2}\right) = 0
 \label{e:Psi:zero}
\end{eqnarray}
and for convenience, we define effective flux function
$\tilde{\Psi}$ as a flux function $\Psi$ multiplied by circumference
$2\pi{r}$ and scaled by the ideal CCF torque ${\cal{T}}_C$
(\ref{e:TC})
\begin{eqnarray}
\tilde{\Psi}(r,z) = \frac{2 \pi r}{{\cal{T}}_C} \Psi(r,z)
 \label{e:Psi:tilde}
\end{eqnarray}
Finally, due to multiplication of flux function $\Psi$ by $r$, the
difference in effective flux function $\tilde{\Psi}$ between two
points gives the total flux of angular momentum across the {\it
conical} \ or {\it cylindrical surface} on which these points lie.
Thus, in the case of e.g.~inner and outer cylinder, it can be shown
that torques ${\cal T}_1$ and ${\cal T}_2$ scaled by the ideal CCF
torque ${\cal{T}}_C$ are given by
\begin{eqnarray}
\frac{{\cal T}_1}{{\cal{T}}_C} = \tilde{\Psi}(R_1,H) -
\tilde{\Psi}(R_1,0) \qquad \frac{{-\cal T}_2}{{\cal{T}}_C} =
\tilde{\Psi}(R_2,H) - \tilde{\Psi}(R_2,0)
 \label{e:Psi:T1}
\end{eqnarray}
Similar expressions hold for the torques applied to the horizontal
surfaces.

\end{document}